\documentclass[11pt]{article}    

\usepackage[margin=1 in]{geometry}% For nicer margins
\usepackage{comment}                % For commenting large portions of text
\usepackage[tbtags]{amsmath}        % Basic math package (includes align environment)
\usepackage{xcolor}                 % For colour (better than the color package)
\usepackage{enumitem}               % For lists (better than the enumerate package)
\usepackage{amssymb}                % Math symbols (includes boldface \mathbb{})
\usepackage{amsthm}                 % Enhances the \newtheorem command (includes the proof environment)
\usepackage{hyperref}               % For referencing (e.g. URLs)
\usepackage{graphicx}               % For including graphics (better than the graphics package)
\usepackage{tikz, pgfplots}         % TiKz is for drawing pictures, pgfplots makes some drawings easier
\usepackage{float}                  % Helpful for positioning figures
\usepackage{natbib}                 % Package used for easy citation
\usepackage{rotating}
\usepackage{setspace}
\usepackage{booktabs}
\usepackage{cleveref}
\usepackage{tcolorbox}
\usepackage{pdfpages}
\usepackage{natbib}

\theoremstyle{theorem}

\newtheorem{proposition}{Proposition}
\newtheorem{lemma}{Lemma}

\numberwithin{equation}{section}

\title{% 
	Nonlinear Forecast Error Variance Decompositions with Hermite Polynomials\thanks{Many thanks to Martin Burda, Antoine Djogbenou, Maksim Isakin, Christian Gourieroux, Yao Luo, Angelo Melino, Kerem Tuzcuoglu, Yuanyuan Wan, the participants of the 59th Annual Meeting of the Canadian Economics Association, the Celebrating James Mackinnon Conference, and the 2025 RCEA International Conference in Economics, Econometrics, and Finance for their insightful comments.}}
\author{\textsc{Quinlan Lee}\footnote{Ph.D. Candidate, University of Toronto, Department of Economics; email: \textit{qt.lee@mail.utoronto.ca}}}
\date{This Version: \today. }

\begin{document}

		\maketitle
		\begin{abstract}
			\noindent 		
					
			A novel approach to Forecast Error Variance Decompositions (FEVD) in nonlinear Structural Vector Autoregressive models with Gaussian innovations is proposed, called the Hermite FEVD (HFEVD). This method employs a Hermite polynomial expansion to approximate the future trajectory of a nonlinear process. The orthogonality of Hermite polynomials under the Gaussian density facilitates the construction of the decomposition, providing a separation of shock effects by time horizon, by components of the structural innovation and by degree of nonlinearity. A link between the HFEVD and nonlinear Impulse Response Functions is established and distinguishes between marginal and interaction contributions of shocks. Simulation results from standard nonlinear models are provided as illustrations and an application to fiscal policy shocks is examined. \\
			
			\vspace{0in} 
		\end{abstract}

	\setstretch{1}
	
	\newpage
	
\section{Introduction}

The Forecast Error Variance Decomposition (FEVD) is a popular tool for modern empirical macroeconomic research. It allows a practitioner to assess the importance of shocks in a dynamic model by partitioning the contributions of its structural innovations on the conditional variance of its forecast. In the linear (Gaussian) Structural Vector Autoregressive (SVAR) framework, the construction of the FEVD is facilitated by the partial moving average representation. Indeed, the future trajectory of a stationary SVAR model can always be written as a sum of its independent structural innovations. Hence, the conditional variance of this sum is additively separable, and it partitions the effects of shocks by time horizon and by components of the structural innovations. \\

Although the assumption of linearity may be appealing in practice, it can oversimplify economic phenomena, leading to misguided conclusions for policymakers. Despite recent advancements in nonlinear modeling, extensions of the FEVD in these contexts remain largely unexplored. To date, only two methods have emerged in the current literature. The first method, proposed by \cite{LN2016}, introduces the Generalized FEVD (GFEVD), based on the Generalized Impulse Response Function (GIRF) developed by \cite{KPP1996}. The second method, introduced by \citet{IN2020}, is a FEVD based on the Law of Total Variance. While these methods have steered the nonlinear FEVD literature into a promising direction, a clear consensus among practitioners is still lacking [\cite{KL2017}, Section 18.2.2].\\

% Moreover, these studies are only interested in the decomposition by structural innovation components, but do not account for nonlinear dynamics adequately.   \\ 

This paper fulfills the gap in the literature and proposes a new method for nonlinear FEVD analysis known as the Hermite FEVD (HFEVD) approach. It can be applied directly to parametric nonlinear SVAR models written on standard Gaussian innovations and to semiparametric nonlinear SVAR models with innovations that are standardized to Gaussian\footnote{While this paper works only with nonlinear SVAR models, the results can be extended to a more general class of nonlinear dynamic models.}. The strategy is to approximate the trajectory of these models using a Hermite polynomial expansion, resulting in a partial Volterra representation of the process at any future horizon. The orthogonality of Hermite polynomials under the standard normal density facilitates the additive separability of the conditional variance for this expansion, leading to the HFEVD. Moreover, its validity is supported by the fact that each term in the HFEVD is directly linked to the impact multipliers of the nonlinear Expected Impulse Response Function (EIRF)\footnote{The proposed methodology can also be generalized to models with non-Gaussian innovations by constructing orthogonal polynomials with respect to their densities, but the link to EIRFs is preserved only with Gaussian innovations and Hermite polynomials.}. This extends an analogous interpretation for the terms in the classical FEVD for the linear SVAR framework.\\

The proposed approach is unique in that it partitions the variance of the forecast along three dimensions: by time horizon, by the components of structural innovations, and by degree of nonlinearity. The latter dimension is particularly useful for practitioners, as it allows assessment of the nonlinearity in shock transmission channels within dynamic models. Moreover, because each term in the HFEVD is orthogonal to one another, they can be computed independently, allowing for a partial decomposition of contributions even when a trajectory is highly nonlinear. The HFEVD also accounts for the state of the economy, which potentially yields a different decomposition depending on the historical context it conditions upon. A practical exposition of these ideas are studied through the lens of the \cite{FRF2015} Threshold VAR framework. \\

The structure of this paper is as follows: Section 2 provides a review of the FEVD and it properties within a linear SVAR framework. Section 3 introduces the nonlinear SVAR model and its future trajectory. Section 4 proposes the Hermite Forecast Error Variance Decomposition and its properties. Section 5 provides details on numerical implementation of the HFEVD and Section 6 offers simulation results. Section 7 features the application of the HFEVD to the FRF2015 TVAR model and Section 8 concludes. Proofs, technical details, and the generalization of the method to non-Gaussian innovations are confined to the Appendix.

\section{FEVD in a Linear Dynamic Framework}

This section reviews the construction of the FEVD and its properties through the lens of the linear Gaussian Structural Vector Autoregressive model. The FEVD is readily available in closed form and is computed easily in this framework. The SVAR(1) model will be introduced alongside its moving average representation, detailing how the trajectory at horizon $h$ can be derived through recursive substitution. Subsequently, the discussion will address the FEVD and its identification challenges. Finally, the relationship between FEVD and Impulse Response Functions (IRFs) will be established.

\subsection{The Linear Gaussian SVAR Model}

Consider an $n$-dimensional linear Gaussian SVAR(1) process $(Y_t)$ given by: 
\begin{equation}\label{SVAR}
	Y_t = AY_{t-1} + D\varepsilon_t,
\end{equation}
where $A$ is an $n \times n$ matrix of autoregressive coefficients, $D$ is an invertible $n \times n$ structural weighting matrix, $(\varepsilon_t)$ are the structural innovations assumed to be independent multivariate standard normal and $\varepsilon_t$ is independent of $y_{t-1}$. When the eigenvalues of $A$ have a modulus strictly less than 1, the system \eqref{SVAR} has a unique stationary solution with an infinite moving average representation of the form:
\begin{equation}\label{wold}
	Y_t = \sum_{i=0}^\infty \Theta_i\varepsilon_{t-i},
\end{equation}
where $\Theta_i = A^iD$ are the coefficient matrices of the moving average terms.  Under the assumption of Gaussian innovations, only the matrix coefficients $A$ and $\Sigma= DD'$ are identifiable in this context, but not $D$ itself \footnote{More precisely, $D$ is identified up to an orthogonal transformation and the degree to which it is underidentified is $\frac{n(n-1)}{2}$. }. Hence, the structural innovations $\varepsilon_t$ and its components are not identifiable without imposing further assumptions on the matrix $D$\footnote{The most straightforward approach would be to order the components of $Y_t$ and assume the matrix $D$ to be lower triangular. However, this approach proposed by \cite{S1980} depends on the ordering and has no real economic interpretation in general. More complex identification methods are also available and are comprehensively reviewed in \cite{R2016}.}.  It is explicitly presumed that the practitioner has applied a sufficient and reasonable set of assumptions to resolve this identification issue such that the shocks on the structural innovations $\varepsilon_t$ have valid economic meaning.\\

The trajectory of the SVAR(1) process at horizon $h$ can be computed by means of recursion. Beginning at time $t+1$, we have: 
\begin{equation*}
	\begin{split}
		Y_{t+1} & = AY_{t} + D\varepsilon_{t+1} \\
		= & AY_{t}+\Theta_0\varepsilon_{t+1}\\
	\end{split}
\end{equation*}
Then, by recursion, at any horizon $h$, the formula becomes:
\begin{equation}\label{pwold}
	\begin{split}
		Y_{t+h} & = A^hY_t + \sum_{i=0}^{h-1} A^{i}D\varepsilon_{t+h-i}= A^hY_t + \sum_{i=0}^{h-1} \Theta_i\varepsilon_{t+h-i}.
	\end{split}
\end{equation}
Thus, a partial moving average representation is obtained at horizon $h$, where  the trajectory $Y_{t+h}$ is partitioned into two parts: [1] The observed history, $A^hY_t$, which contains the weighted path of all past innovations. [2] The weighted path of future innovations, $\sum_{i=0}^{h-1} \Theta_i\varepsilon_{t+h-i}$. By conditioning on $Y_t$, the trajectory of this process is stochastic only through the future innovations $\varepsilon_{t+i}$, for $i=1,...,h$. Hence, a variance decomposition in this context aims to separate the effects of these innovations on this stochastic component. 

\subsection{Forecast Error Variance Decomposition (FEVD)}

The FEVD is a partition of the (conditional) variance covariance matrix for the future trajectory. For instance, a decomposition by time (innovation) horizon is given by: 
\begin{equation}\label{hfevdsvar}
	\mathbb{V}\left[Y_{t+h}|Y_t\right] = \sum_{i=0}^{h-1} A^{i}DD'(A^{i})'=\sum_{i=0}^{h-1} \Theta_i \left(\Theta_i\right)', 
\end{equation}
where $\Theta_i=A^{i-1}D$ is called the multiplier matrix. Each term in the decomposition is the contribution of the structural innovation at the $i$-th horizon, $\varepsilon_{t+i}$ \footnote{Another interpretation of equation \eqref{hfevdsvar} is the term structure of total risk arising from the forecast over $h$ horizons. Each component on the right hand side corresponds to the updating of information sets between time $t+i-1$ and $t+i$, and measures the risk along the ``maturity horizon". More generally, for any square integrable process $(Y_t)$ with information set $(I_t)$, the decomposition by maturity horizon is given by: $\mathbb{V}[Y_{t+h}|I_t]=\sum_{i=1}^{h} \mathbb{V}\{[\mathbb{E}(Y_{t+h}|I_{t+i})-\mathbb{E}(Y_{t+h}|I_{t+i-1})]|I_t\}$ [see \cite{GL2024}, Lemma 2]. In the linear framework, the notions of decomposition by ``innovation" horizon and ``maturity" horizon are the same, but this will not be the case in the nonlinear framework. In this particular, this paper is focused on the innovation horizon which is of interest to macroeconomists.}. Notably, the horizonal decomposition is a function of $A$ and $DD'$, so it is always identifiable. The decomposition in \eqref{hfevdsvar} is also independent of the conditioning value $Y_t$. This is known as the history invariance property of the FEVD in the linear dynamic model; regardless of what we have seen in the most recent time period, the decomposition of risk remains the same. Clearly, this is not a very realistic assumption in practice\footnote{For example, we would expect the decomposition of financial or macroeconomic risk to be different when the interest rates are low compared to when the interest rates are high.}. \\

Another decomposition of interest is the contribution to the variance-covariance matrix of $Y_{t+h}$ accounted for by the individual components of the structural innovations. Let us denote the $j$-th column of the multiplier matrix as $(\Theta_{i})_{*,j}=(A^{i}D)_{*,j}$ (an $n \times 1$ vector). Then we have: 
\begin{equation}\label{hefevdsvar}
	\mathbb{V}\left[Y_{t+h}|Y_t\right] = \sum_{i=0}^{h-1} \sum_{j=1}^n (\Theta_{i})_{*,j}(\Theta_{i})_{*,j}'.
\end{equation}
The product term $(\Theta_{i})_{*,j}(\Theta_{i})_{*,j}'$ captures the effect of the $j$-th structural innovation at horizon $t+i$ on the trajectory of the process. For instance, the $p,q$-th entry in $(\Theta_{i})_{*,j}(\Theta_{i})_{*,j}'$ corresponds to the contribution of $\varepsilon_{j,t+i}$ on the covariance between $Y_{p,t+i}$ and $Y_{q,t+i}$. Hence, the FEVD in \eqref{hefevdsvar} now provides a decomposition by time horizon and by structural innovation component. However $(\Theta_{i})_{*,j}$ depends on the parameter $D$ itself, so the decomposition by components is not identifiable without further restrictions on $D$. \\

In general, a practitioner is more interested in the scalar decomposition of some component of $Y_{t+h}$ or their linear combinations. For instance, the horizonal decomposition in \eqref{hfevdsvar} for a linear combination $\alpha' Y_{t+h}$ can be written as:
\begin{equation}\label{vdl1}
	\mathbb{V}[\alpha' Y_{t+h}|Y_t] =\sum_{i=0}^{h-1} \alpha' \Theta_i \left(\Theta_i\right)'\alpha,
\end{equation}
or a decomposition by components in \eqref{hefevdsvar} as: 
\begin{equation}\label{vdl2}
	\mathbb{V}[\alpha' Y_{t+h}|Y_t] =\sum_{i=0}^{h-1}\sum_{j=1}^n \left[\alpha'(\Theta_{i})_{*,j}\right]^2.
\end{equation}
In the special case where $\alpha$ is a vector equal to 1 in the $\ell$-th entry and $0$, otherwise, the FEVD is a decomposition of the conditional variance for component $Y_{\ell,t+h}$:
\begin{equation}
	\mathbb{V}[Y_{\ell,t+h}|Y_t]= \sum_{i=0}^{h-1} \sum_{j=1}^n \left[(\Theta_{i})_{\ell,j}\right]^2.
\end{equation}

\subsection{Impulse Response Functions (IRF)}

An important property of the FEVD in the linear framework is its relationship to Impulse Response Functions. Intuitively, it evaluates the consequence of a transitory shock to the structural innovation at date $t+1$ by comparing two trajectories of $Y_{t+h}$: [1] A perturbed path, where a shock magnitude of $\delta$ is applied to structural innovation $\varepsilon_{t+1}$. [2] A baseline path, where the process follows its natural trajectory defined in \eqref{pwold}. To formalize this idea, suppose the practitioner is interested in a shock of magnitude $\delta$ at time period $t+1$. The perturbed and baseline paths at horizon $h$ are given by: 
\begin{equation}
	\begin{split}
		Y_{t+h}(\delta) = & A^h Y_t + A^{h-1}D(\varepsilon_{t+1}+\delta) +  \sum_{i=0}^{h-2}A^{i}D\varepsilon_{t+h-i}, \\
		Y_{t+h}& = A^h Y_t + \sum_{i=0}^{h-1}A^{i}D\varepsilon_{t+h-i}.  \\
	\end{split}
\end{equation}
To evaluate the consequence of this conceptual experiment, the difference between these two paths is considered: 
\begin{equation}\label{uni_onetime_IRF}
	IRF(h,\delta)=Y_{t+h}(\delta)-Y_{t+h} =  A^{h-1}D\delta=\Theta_{h-1}\delta, \ \forall h, \delta.
\end{equation}
Equation \eqref{uni_onetime_IRF} defines the Impulse Response Function (IRF), which evaluates the effect of the shock through $\Theta_{h-1}=A^{h-1}D=\frac{\partial Y_{t+h}}{\partial \varepsilon_{t+1}}$. This is called the impact multiplier, which captures ``consequences of a one-unit increase in the [structural] innovation" on the trajectory [\cite{H1994}, Definition 11.4.2]. Like the FEVD, the IRF is history invariant in the linear framework, since it is independent of the most recent value $Y_t$. There is a drift effect at each horizon, which is linear in the magnitude vector $\delta$. We also note that the IRF does not depend on the path of future innovations, and coincides with the (conditional) expected IRF, that is:
\begin{equation}
	EIRF(h,\delta,Y_t) = \mathbb{E}\left[Y_{t+h}(\delta)-Y_{t+h}|Y_t\right] = IRF(h,\delta).
\end{equation}
It follows from equation \eqref{uni_onetime_IRF} that the FEVD can be written as:
\begin{equation}\label{lfevdirf}
	\mathbb{V}\left[Y_{t+h}|Y_t\right] = \sum_{i=1}^h \sum_{j=1}^n IRF(i,e_j) IRF(i,e_j)',
\end{equation}
where $e_j$ is a vector of 0's except for the $j$-th entry which is equal to 1. Intuitively, the FEVD is a sum of squared IRFs following shocks of unit magnitude, where the term $IRF(i,e_j)$ corresponds to the effect of a unit shock on the element $\varepsilon_{j,t+1}$ at horizon $i$.

\section{The Nonlinear SVAR Model}

This section introduces the Nonlinear SVAR process and explains how the future trajectory can be imputed by means of recursive substitution. Some examples of nonlinear data generating processes that fall under this framework are also provided.

\subsection{Nonlinear SVAR(p) Process}

First, a general class of nonlinear models, broadly referred to as the Nonlinear SVAR process, can be introduced through the following proposition:
\begin{proposition}
	$(Y_t)$ is a Markov process of order $p$ with a strictly positive transition density if and only if it admits a nonlinear Structural Vector Autoregressive representation of the form:
	\begin{equation}\label{NLSVAR}
		Y_t = g(\underline{Y_{t-1}};\varepsilon_t), t \geq 1,
	\end{equation}
	where the vector $\varepsilon_t=(\varepsilon_{1,t},...,\varepsilon_{n,t})'$ is standard multivariate Gaussian with covariance matrix $I_d$, $\varepsilon_t$ is independent of the history $\underline{Y_{t-1}}=(Y_{t-1},...,Y_{t-p})$ and $g$ is an invertible function with respect to $\varepsilon_t$.
\end{proposition}

\noindent \textbf{Proof:} This is a direct consequence of the triangular representation of joint continuous distributions [\cite{R1952}]. \qed \\

The nonlinear structural Gaussian innovations $(\varepsilon_t)$ are given by:
\begin{equation}\label{nlminno}
	\varepsilon_t = g^{-1}(Y_t,\underline{Y_{t-1}}),
\end{equation}
where $g^{-1}(\cdot,\underline{Y_{t-1}})$ denotes the inverse of the function $g(\underline{Y_{t-1}},\cdot)$. These innovations are nonlinear functions of $Y_t$ given its past, and are the basis for structural shocks in the construction of IRFs and FEVDs which will be defined below. There are a number of nonlinear models in the literature which fall under this paradigm\footnote{For a comprehensive review of such models, see \cite{KL2017} Chapter 19.}, including but not limited to Threshold and Smooth Transitions [e.g. \cite{HT2013}, \cite{GM2014}, \cite{FRF2015}], Time-Varying Parameter [e.g. \cite{P2005}, \cite{N2011}] and Nonparametric SVARs [e.g.\cite{HTY1998}, \cite{J2013}]. \\

In general, the function $g$ and its structural innovations $(\varepsilon_t)$ are not identifiable unless the dimension of $(Y_t)$ is equal to 1. This problem is analogous to the identification issue of the matrix $D$ in the linear Gaussian SVAR framework \eqref{SVAR}. Depending on the nature of nonlinearity selected and the information available to the practitioner, this identification issue can be resolved in different ways\footnote{For example, for nonlinear SVAR models with additive errors, one may consider Independent Component Analysis (ICA) as a solution to the identification issue, if at most one component of the structural innovations is Gaussian [see \cite{GMR2017} for a discussion and review on ICA in SVAR models]. In such a context, the non-Gaussian innovations will have to be re-normalized inT order to satisfy the representation \eqref{NLSVAR} as discussed below.}. However, to avoid detracting from the main ideas in this paper, it is explicitly assumed that the practitioner has selected an appropriate identification scheme with discretion, so that a shock on the structural Gaussian innovations $(\varepsilon_t)$ is indeed compatible with plausible economic interpretations.  \\

\noindent \textbf{Remark 1:} The nonlinear representation in \eqref{NLSVAR} is not unique to Gaussian innovations. More specifically, a Markov process $(Y_t)$ may also be written as: 
\begin{equation}\label{NLAR_alt}
	Y_t = g(\underline{Y}_{t-1},\varepsilon_t)= g^*(\underline{Y}_{t-1};u_t),
\end{equation}
where $g^*$ is invertible in $u_t=(u_{1,t},...,u_{n,t})$, $u_t$ has independent components and each component has a conditional  c.d.f. $u_{j,t}\sim F_j$ for $j=1,...,n$., which are not necessarily Gaussian. Proposition 1 can still be used by changing the sign of $u_t$ (to transform a decreasing function to an increasing one) and transforming the components of $u_t$ by $\Phi^{-1}F_j(\cdot)$ for $j=1,...,n$, where $\Phi$ denotes the standard normal c.d.f. (to transform $F_j$ into the Gaussian). The choice of a representation with Gaussian innovations is simply a normalization\footnote{While the main text will provide methodology and the key results for models with Gaussian innovations, Appendix D will extend the ideas for representations with non-Gaussian errors.}, in line with the convention set by the literature [see \cite{L2019}, Section 6.3 for a discussion of normalization in identification], and to facilitate a link between IRFs and FEVDs in the nonlinear framework [see the discussion in Section 4.4]. 

\subsection{Future Trajectories}

Beginning at time $t$, we may recursively define the trajectory of the process at future horizons using recursive substitution. Let us denote the relevant history at time $t$ as $\underline{Y}_{t} = (Y_{t},...,Y_{t-p+1})$. Based on \eqref{NLSVAR}, the process $(Y_{t})$ is such that:
\begin{equation*}
	Y_{t+1} = g(\underline{Y_t},\varepsilon_{t+1}).
\end{equation*}
Applying the recursive formula at $h=2$, we obtain: 
\begin{equation}
	\begin{split}
		Y_{t+2} & = g(Y_{t+1},...,Y_{t-p+2},\varepsilon_{t+1})\\
		& = g(g(\underline{Y_t},\varepsilon_{t+1}),\varepsilon_{t+2}).
	\end{split}
\end{equation}
In general, for a horizon $h$, this substitution procedure yields:
\begin{equation}\label{pvolt}
	Y_{t+h}  = g^{(h)}(\underline{Y_t},\varepsilon_{t+1:t+h}), 
\end{equation}
where $\varepsilon_{t+1:t+h}=(\varepsilon_{t+1},...,\varepsilon_{t+h})'$ denotes the future path of structural innovations between $t+1$ and $t+h$, and $g^{(h)}$ represents the iteration of function $g$ a total of $h$ times. Thus the value of the process at time $t+h$ is a function of the relevant history $\underline{Y}_{t}$ and the sequence of all future structural innovations between $t+1$ and $t+h$. For relatively simple functions $g$, it is possible to obtain $Y_{t+h}$ in closed form. However, this may be challenging when the process (and the function $g$) is highly nonlinear. In the latter case, this trajectory can be obtained by means of simulation (see Section 5 for a discussion).

\subsection{Examples}

\subsubsection{Double Autoregressive Model (DAR)}

The univariate Double Autoregressive model of order one [\cite{L2007}], denoted DAR(1), is used to model conditional heteroscedasticity. It is given by the representation: 
\begin{equation}\label{DAR}
	y_t = \phi y_{t-1} + \sqrt{\alpha + \beta y_{t-1}^2} \varepsilon_t \equiv g(y_{t-1},\varepsilon_t),
\end{equation}
where $\varepsilon_t$ is iid N(0,1), $\alpha >0$, and $\beta \geq 0$\footnote{Moreover, the model admits a strictly stationary solution if $\mathbb{E}\left[\log|\phi + \sqrt{\beta}\varepsilon_t|\right]<0$.}.   Therefore, function $g$ is nonlinear in $y_{t-1}$, linear in $\varepsilon_t$, with nonlinear cross effects between $y_{t-1}$ and $\varepsilon_{t}$. Its Gaussian innovation is given by the expression: 
\begin{equation}
	\varepsilon_t = \frac{y_t - \phi y_{t-1}}{\sqrt{\alpha + \beta y_{t-1}^2}}\equiv g^{-1}(y_t,y_{t-1}) .
\end{equation}
Through recursive substitution, we get the following representation for $h=2$:
\begin{equation}\label{dar2}
	\begin{split}
		y_{t+2}  = & \phi y_{t+1} + \sqrt{\alpha + \beta y_{t+1}^2} \varepsilon_{t+2}\\
		= & \phi^2 y_{t} + \phi\sqrt{\alpha + \beta y_{t}^2} \varepsilon_{t+1}  + \sqrt{\alpha + \beta \left(\phi y_{t} + \sqrt{\alpha + \beta y_{t}^2} \varepsilon_{t+1} \right)^2} \varepsilon_{t+2}. \\
	\end{split}
\end{equation}
Even at horizon $h=2$, the expression \eqref{dar2} is quite complicated, which features interaction terms between innovations at different horizons. By continuing the substitution process, it would be difficult to obtain a general formula for $y_{t+h}$, since there would be a compounding of terms from the previous horizons. \\

\noindent \textbf{Remark 2:} It is also possible to consider a semi-parametric version of Example 1 of the form: 
\begin{equation}\label{DARa}
	y_t = \phi y_{t-1} + \sqrt{\alpha + \beta y_{t-1}^2} u_t,
\end{equation}
where the innovation process $u_t$ is a strong white noise process, not necessarily Gaussian. Let $F$ denote its corresponding cumulative distribution function of $u_t$ and $\Phi$ denote the standard normal cumulative distribution function. Then, the model \eqref{DARa} can be written as:
\begin{equation}\label{DARb}
	y_t = \phi y_{t-1} + \sqrt{\alpha + \beta y_{t-1}^2}\left[F^{-1}(\Phi(\varepsilon_t))\right],
\end{equation}
where the nonlinear autoregression \eqref{DARa} is now written as a function of the Gaussian innovations $\varepsilon_t \sim N(0,1)$. In this extension, the function $g$ is also nonlinear in $\varepsilon_t$. The model is semi-parametric with scalar parameters $\phi$, $\alpha$, $\beta$ and functional parameter $F$.  

\subsubsection{Stochastic Volatility with Regimes} 

Consider a discrete time stochastic volatility model where the daily return $(y_t)$ and the within-day logged realized volatility $(z_t)$, i.e. the logged variance of 5 minute returns in day $t$, evolve according to the model: 
\begin{equation}\label{threshold}
	\begin{split}
		y_t& = a + by_{t-1} + \left[a_1\times\textbf{1}_{y_{t-1}\geq0} + a_2 \times \textbf{1}_{y_{t-1}<0}\right]z_{t-1}+\exp(z_{t-1})\varepsilon_{1,t}, \\
		z_t& = \phi z_{t-1} + \varepsilon_{2,t},\\
	\end{split}
\end{equation}
where $\varepsilon_t=(\varepsilon_{1,t},\varepsilon_{2,t})'$ is a bivariate standard Gaussian innovation. There are two nonlinearities introduced in the autoregressive representation: [1] The endogenous threshold transition between the regimes $y_{t-1} \geq 0$ and $y_{t-1}<0$. [2] The interaction between $\exp(z_{t-1})$ and $\varepsilon_{1,t}$. For exposition, let us compute the trajectory of this process at horizon $h=1$:
\begin{equation}
\begin{split}
	y_{t+1} & = a + by_t + \left[a_1\times\textbf{1}_{y_{t}\geq0} + a_2 \times \textbf{1}_{y_{t}<0}\right]z_{t}+\exp(z_{t})\varepsilon_{1,t+1}\\ 
	& = \left[a + by_t +a_1z_t + \exp(z_t)\varepsilon_{1,t+1}\right](\textbf{1}_{y_{t}\geq0}) + \left[a + by_t +a_2z_t + \exp(z_t)\varepsilon_{1,t+1}\right](\textbf{1}_{y_{t}<0}).
\end{split}
\end{equation}
The conditioning value (history) $y_t$ plays an important role on this trajectory - in particular, it influences the regime of $y_{t+1}$. Similarly, by recursive substitution, we obtain for horizon $h=2$:
\begin{equation}
	\begin{split}
		y_{t+2} & = \left[a+ba+b^2y_t + ba_1z_t + b\exp(z_t)\varepsilon_{1,t+1}+a_1(\phi z_{t}+\varepsilon_{2,t+1})+\exp(\phi z_{t}+\varepsilon_{2,t+1})\varepsilon_{1,t+2}\right]\\
		& \times (\textbf{1}_{y_{t}\geq0})\times (\textbf{1}_{a + by_t +a_1z_t + \exp(z_t)\varepsilon_{1,t+1}\geq0}) \\ 
		& + \left[a+ba+b^2y_t + ba_1z_t + b\exp(z_t)\varepsilon_{1,t+1}+a_2(\phi z_{t}+\varepsilon_{2,t+1})+\exp(\phi z_{t}+\varepsilon_{2,t+1})\varepsilon_{1,t+2}\right]\\
		& \times (\textbf{1}_{y_{t}\geq0})\times (\textbf{1}_{a + by_t +a_1z_t + \exp(z_t)\varepsilon_{1,t+1}<0}) \\ 
	    & + \left[a+ba+b^2y_t + ba_2z_t + b\exp(z_t)\varepsilon_{1,t+1}+a_1(\phi z_{t}+\varepsilon_{2,t+1})+\exp(\phi z_{t}+\varepsilon_{2,t+1})\varepsilon_{1,t+2}\right]\\	
		& \times (\textbf{1}_{y_{t}<0})\times (\textbf{1}_{a + by_t +a_2z_t + \exp(z_t)\varepsilon_{1,t+1}\geq0}) \\ 
		& + \left[a+ba+b^2y_t + ba_2z_t + b\exp(z_t)\varepsilon_{1,t+1}+a_2(\phi z_{t}+\varepsilon_{2,t+1})+\exp(\phi z_{t}+\varepsilon_{2,t+1})\varepsilon_{1,t+2}\right]\\
		& \times (\textbf{1}_{y_{t}\geq0})\times (\textbf{1}_{a + by_t +a_2z_t + \exp(z_t)\varepsilon_{1,t+1}\geq0}), \\ 
	\end{split}
\end{equation}
$y_{t+2}$ is function $y_t$, $z_t$ and  the future innovations $\varepsilon_{1,t+1}$, $\varepsilon_{1,t+2}$ and $\varepsilon_{2,t+1}$. There are four regimes, which depend on the values of $y_t$ and $y_{t+1}$. In general, $2^h$ such regimes exist at horizon $h$, and the trajectory of a threshold process can be viewed as a ``tree" of nodes that depend on the evolution of the threshold variable throughout the time horizons. As such, when analyzing shocks in a threshold framework, it is important to take into account the fact that there can be a switching of regimes between horizons\footnote{In particular, by conditioning on $y_t$, you may govern in which regime the process starts, but you cannot control whether or not it stays in that regime, since it depends on the future innovations $\varepsilon_{t+i}=(\varepsilon_{1,t+i},\varepsilon_{2,t+i})$ for $i=1,...,h$.}. \\

\noindent \textbf{Remark 3:} A smoothed version of \eqref{threshold} is given by: 
\begin{equation}\label{smootht}
	\begin{split}
		y_t& = a +  by_{t-1}+\left[a_1+\frac{a_2-a_1}{1+\exp\left(-y_{t-1}\right)}\right]z_{t-1}   + \exp(z_{t-1})\varepsilon_{1,t}, \\
		z_t& = \phi z_{t-1} + \varepsilon_{2,t},\\
	\end{split}
\end{equation}
where a transition, characterized by a first-order logistic function, governs the evolution of the risk premium coefficients.

\section{Hermite Forecast Error Variance Decomposition (HFEVD)}

In the linear framework, recursive substitution yields directly a partial moving average representation \eqref{pwold}. However, when nonlinearities are present, this is no longer the case, and there is no reason for the expression obtained in \eqref{pvolt} to have a ``nice" closed form expression that can facilitate the construction of a valid FEVD. The approach in this paper, referred to hereafter as the Hermite Forecast Error Variance Decomposition (HFEVD), utilizes a Hermite polynomial expansion to approximate the future trajectory\footnote{Recall from Remark 1 that the nonlinear SVAR can be written on innovations that are non-Gaussian. Indeed, the ideas behind the HFEVD can be generalized to other sequences of orthogonal polynomials. However, not all distributions are associated with a natural set of orthogonal polynomials. Appendix D.1 will discuss how orthogonal polynomials for an arbitrary distribution may be constructed, and Appendix D.2 will extend the HFEVD approach for non-Gaussian errors.}. The resulting nonlinear representation is a sum of uncorrelated terms, which allows us to separate the variance along three dimensions: by the time horizon, by the components of the structural innovations and by the degree of nonlinearity. 

\subsection{Hermite Polynomials}

\subsubsection{Definition and Properties}

A sequence of polynomials of degree $k$, denoted $f_k(\varepsilon)$ for $\varepsilon \in \mathbb{R}$, are orthogonal on the interval $(-\infty,\infty)$, with respect to the weight function $w(\varepsilon)$ if they satisfy:
\begin{equation}\label{ortho_poly}
	\int_{-\infty}^{\infty} f_k(x)f_m(x)w(x)dx = 0,
\end{equation}
where $k\neq m$ and $k,m \in \mathbb{N}$ [\cite{AS1964}]. A special case of orthogonal polynomials are the Hermite polynomials, where the weight function $w(x)=\frac{1}{\sqrt{2\pi}}\exp[-x^2/2]$ coincides with the standard normal density. In the univariate case, the $k$-th Hermite polynomial is given by: 
\begin{equation}\label{hermite_uni}
	H_k(\varepsilon)=(-1)^k\exp\left(\frac{\varepsilon^2}{2}\right)\frac{d^k}{d\varepsilon^k}\left[\exp\left(-\frac{\varepsilon^2}{2}\right)\right],
\end{equation}
for $k \in \mathbb{N}$. Beginning with the initial value $H_0(\varepsilon)=1$, it follows that the sequence of Hermite polynomials can be defined by the recursive relation: 
\begin{equation}\label{hermitechain}
	H_{k+1}(\varepsilon) = 2\varepsilon H_k(\varepsilon) - \frac{dH_k(\varepsilon)}{d\varepsilon},
\end{equation}
which is a consequence of applying the chain rule directly on the definition \eqref{hermite_uni}. For example, the first six Hermite polynomials are:
\begin{table}[h!]
	\centering
	\begin{tabular}{lll}
		$H_0(\varepsilon) =1$&$H_2(\varepsilon) =\varepsilon^2-1$ & $H_4(\varepsilon) =\varepsilon^4-6\varepsilon^2+3$  \\
		$H_1(\varepsilon) =\varepsilon$&$H_3(\varepsilon) =\varepsilon^3-2\varepsilon$ & $H_5(x) =\varepsilon^5-10\varepsilon^3+15\varepsilon$. \\
	\end{tabular}
\end{table}

This paper works exclusively with Hermite polynomials of multivariate normal random variables. Suppose  $\varepsilon=(\varepsilon_1,...,\varepsilon_h)'$ such that $\varepsilon \sim MVN(0,Id)$. Then, the joint Hermite polynomial of $\varepsilon$ is equal to the product of the univariate Hermite polynomials for each component in $\varepsilon$:
\begin{equation}\label{multihermite}
	H_K(\varepsilon) = \prod_{i=1}^h H_{k_i}(\varepsilon_i),
\end{equation}
where $k_i$ represents the degree of the marginal Hermite polynomials for $\varepsilon_i$, $1 \leq i \leq h$, and $K=(k_1,...,k_h)'$ is a vector in $\mathbb{N}^h$. These polynomials have the following first and second-order moments: 
\begin{lemma}
Suppose $\varepsilon=(\varepsilon_1,...,\varepsilon_h)'\sim MVN(0,Id)$ and $H_K(\varepsilon)$ denote a joint Hermite polynomial, where $K=(k_1,...,k_h)' \in \mathbb{N}^h$. Then:  
	\begin{enumerate}
	\item $\mathbb{E}\left[H_K(\varepsilon)H_M(\varepsilon)\right]=0 , \ \ \forall K \neq M. $
		\item $\mathbb{E}\left[H_K(\varepsilon)\right]= \begin{cases}
			1, \ \text{for $K=\textbf{0}$,} \\
			0, \ \text{otherwise.} \\
		\end{cases}$ 
		\item 	$\mathbb{V}\left[H_K(\varepsilon)\right]= \prod_{i=1}^{h}(k_i!)^2, \ \forall K \neq \textbf{0}.$ 
	\end{enumerate}
\end{lemma}

\noindent \textbf{Proof:}  Property (1) follows by definition of orthogonality under the standard normal density. Property (2) follows directly from Property (1) since $\mathbb{E}\left[H_K(\varepsilon)\right]=\mathbb{E}\left[H_K(\varepsilon)H_0(\varepsilon)\right]=0$. Lastly, Property (3) is a special case of \cite{R2017}, Proposition 8, for independent Gaussian random variables.  \qed  

\subsubsection{Approximating Functions of Gaussian Random Variables}

The joint Hermite polynomials form an orthogonal basis of $L^2(\mathbb{R}^h,\mathcal{B},MVN(0,Id))$ ($\mathcal{B}$ denotes the Borel set of $\mathbb{R}^h$), that is, the Hilbert space of square integrable functions with respect to the standard multivariate normal distribution. Hence, any measurable function $g$ of standard Gaussian random variables $\varepsilon=(\varepsilon_1,...,\varepsilon_h)'$ in this space can be approximated by a Hermite polynomial expansion. This is formalized in the following lemma:

\begin{lemma}
	Let $\varepsilon=(\varepsilon_1,...,\varepsilon_h)'$ be a vector of independent standard normal random variables, and $g(\varepsilon)$ be a function in $L^2(\mathbb{R}^h,\mathcal{B},MVN(0,Id))$, that is, square integrable with respect to the (multivariate) standard Gaussian distribution. Then $g(\varepsilon)$ can be written as: 
%	\begin{equation}\label{uni_hermite}
%		g(\varepsilon) = \sum_{k=0}^{\infty}\sum_{\substack{K\in\mathbb{N}^h \\ K:\textbf{1}'K=k}}c_KH_K(\varepsilon),
%	\end{equation}
	\begin{equation}\label{uni_hermite}
			g(\varepsilon) =\sum_{K \in \mathbb{N}^4}c_KH_K(\varepsilon),
		\end{equation}
\end{lemma}
\noindent with $c_\textbf{0} = \mathbb{E}\left[g(\varepsilon)\right]$, $c_K = \mathbb{E}\left[g(\varepsilon)H_K(\varepsilon)\right]/\sqrt{\mathbb{V}(H_K(\varepsilon))}$, $\forall K\neq\textbf{0}$.\\

\noindent \textbf{Proof:} This is a direct application of \cite{R2017}, Corollary 15. \qed \\

Each term in \eqref{uni_hermite} consists of the joint Hermite polynomial $H_K(\varepsilon)$, where $K=(k_1,...,k_h)$ captures the degrees of nonlinearity for the marginal Hermite polynomials of the product, and $c_K$ is the associated coefficient in the Hermite polynomial expansion. The sum is taken across all possibilities in $\mathbb{N}^h$. As an example, consider the first six terms of the expansion for a bivariate standard Gaussian vector $\varepsilon=(\varepsilon_1,\varepsilon_2)'$. We get:
\begin{equation}
	\begin{split}
			g(\varepsilon) = & c_{(0,0)}\times 1+ c_{(1,0)}H_{1}(\varepsilon_1) +  c_{(0,1)}H_{1}(\varepsilon_2) \\
			&  +  c_{(2,0)}H_{2}(\varepsilon_1)  +  c_{(0,2)}H_{2}(\varepsilon_2) + c_{(1,1)}H_{1}(\varepsilon_1)H_{2}(\varepsilon_1) + ... \\
	\end{split}
\end{equation}

The first term corresponds to the initial polynomial $H_{(0,0)}=1$ of degree 0 with coefficient equal to $c_{(0,0)}=\mathbb{E}[g(\varepsilon)]$. The second and third terms correspond to the univariate Hermite polynomials of degree 1 for $\varepsilon_1$ and $\varepsilon_2$, respectively. The fourth and fifth terms are the univariate Hermite polynomials of degree 2 for $\varepsilon_1$ and $\varepsilon_2$, respectively. The sixth term is a joint Hermite polynomial which is a product of the two univariate Hermite polynomials of degree 1 for both $\varepsilon_1$ and $\varepsilon_2$ (and so on).

\subsection{Univariate HFEVD}

As discussed in Sections 3.2 and 3.3, the trajectory of a nonlinear process can be highly complex and may not be available in closed form. As such, it is difficult to obtain a valid decomposition of the conditional variance on \eqref{pvolt} directly. Instead, we may use Lemma 2 to approximate the trajectory using Hermite polynomial expansions to obtain what is known as the partial Volterra representation for the process\footnote{Our approach follows closely to the representation theorem discussed in Section 2.2 of Gourieroux and Jasiak (2005). The objective of their paper is to define nonlinear Impulse Response Functions in the univariate nonlinear framework. In contrast, our goal is to facilitate the construction of a nonlinear FEVD.}; it is the nonlinear analogue to the partial moving average representation in \eqref{pwold} for the linear framework [see \cite{P1978}, Section 4]. The conditional variance on this expansion can be separated additively due to the orthogonality of Hermite polynomials which facilitates the construction of the FEVD. To formalize this idea, let us consider the univariate nonlinear AR(p) case. Its Hermite Forecast Error Variance Decomposition (HFEVD) is given by the following proposition: 

\begin{proposition}[Univariate HFEVD]
%	Suppose the future trajectory of a univariate nonlinear $AR(p)$ at horizon $h$ is given by: 
%	\begin{equation*}
%		y_{t+h} = g^{(h)}(\underline{y_t},\varepsilon_{t+1:t+h}),
%	\end{equation*}
%	where $g^{(h)}$ is square integrable and $\varepsilon_{t+1:t+h}=(\varepsilon_{t+1},...,\varepsilon_{t+h})$ is a sequence of independent standard Gaussian innovations at future horizons. Its variance, conditional on the observed history $\underline{y_t}$, is given by: 
%	\begin{equation}\label{fevd_quin_uni}
%		\mathbb{V}[Y_{t+h}|\underline{y_t}] \approx \sum_{k=1}^{\infty}\sum_{\substack{K\in\mathbb{N}^h \\ K:\textbf{1}'K=k}}\left[c^{(h)}_K(\underline{y_t})\prod_{i=1}^{h}(k_i!)\right]^2, 
%	\end{equation}
%	where $c^{(h)}_K(\underline{y}_t)=\mathbb{E}[g^{(h)}(\underline{y}_t,\varepsilon_{t+1:t+h})H_K(\varepsilon_{t+1:t+h})]/\prod_{i=1}^{h}(k_i!)$ and expectation is taken with respect to the future innovations.
	Suppose the future trajectory of a univariate nonlinear $AR(p)$ at horizon $h$ is given by: 
	\begin{equation*}
			y_{t+h} = g^{(h)}(\underline{y_t},\varepsilon_{t+1:t+h}),
		\end{equation*}
	where $g^{(h)}$ is square integrable and $\varepsilon_{t+1:t+h}=(\varepsilon_{t+1},...,\varepsilon_{t+h})$ is a sequence of independent standard Gaussian innovations at future horizons. Its variance, conditional on the observed history $\underline{y_t}$, is given by: 
	\begin{equation}\label{fevd_quin_uni}
			\mathbb{V}[Y_{t+h}|\underline{y_t}] \approx \sum_{K \in \mathcal{K}}\left[c^{(h)}_K(\underline{y_t})\prod_{i=1}^{h}(k_i!)\right]^2, 
		\end{equation}
	where $\mathcal{K} = \left\{\mathbb{N}^4 \ \backslash \ \textbf{0}\right\}$,  $c^{(h)}_K(\underline{y}_t)=\mathbb{E}[g^{(h)}(\underline{y}_t,\varepsilon_{t+1:t+h})H_K(\varepsilon_{t+1:t+h})]/\prod_{i=1}^{h}(k_i!)$ and expectation is taken with respect to the future innovations.
\end{proposition}

\noindent \textbf{Proof:} From \eqref{pvolt}, the trajectory of a nonlinear AR(p) process is given by:
\begin{equation*}
	y_{t+h} = g^{(h)}(\underline{y}_t,\varepsilon_{t+1:t+h}),
\end{equation*}
where $g^{(h)}$ is square integrable function of the $h$ Gaussian innovations $\varepsilon_{t+1:t+h}=(\varepsilon_{t+1},...,\varepsilon_{t+h})$ at future horizons. Thus, it satisfies the assumptions of Lemma 2, and we get a Hermite polynomial expansion of $y_{t+h}$ as:
	\begin{equation}\label{coro1}
	%\begin{split}
%	y_{t+h} = g^{(h)}(\underline{y}_t,\varepsilon_{t+1:t+h}) = \sum_{k=0}^{\infty}\sum_{\substack{J\in\mathbb{N}^h \\ K:\textbf{1}'K=k}}c^{(h)}_K(\underline{y}_t)H_K(\varepsilon_{t+1:t+h}),
	y_{t+h} = g^{(h)}(\underline{y}_t,\varepsilon_{t+1:t+h}) = \sum_{K \in \mathbb{N}^h}c^{(h)}_K(\underline{y}_t)H_K(\varepsilon_{t+1:t+h}),
	%\end{split}
\end{equation}
with $c^{(h)}_\textbf{0}(\underline{y_t}) = \mathbb{E}\left[g^{(h)}(\underline{y}_t,\varepsilon_{t+1:t+h})\right]$ and $c^{(h)}_K(\underline{y_t}) = \mathbb{E}\left[g^{(h)}(\underline{y}_t,\varepsilon_{t+1:t+h})H_K(\varepsilon_{t+1:t+h})\right]/\sqrt{\mathbb{V}(H_K(\varepsilon_{t+1:t+h}))}$, $\forall K\neq\textbf{0}$. The orthogonality property of Hermite polynomials (Lemma 1(1)) for Gaussian random variables means that the variance of their sum is equal to the sum of their variances. Hence, taking the variance of \eqref{coro1} conditional on $\underline{y_t}$ we get:
	\begin{equation}\label{coro2}
	%\begin{split}
	\mathbb{V}\left[Y_{t+h}|\underline{y_t}\right]\approx  \sum_{K \in \mathcal{K}}\mathbb{V}\left[c^{(h)}_K(\underline{y}_t)H_K(\varepsilon_{t+1:t+h})\big\vert\underline{y_t}\right],
	%\end{split}
\end{equation}
where:
\begin{equation*}
	c^{(h)}_K(\underline{y_t})= \mathbb{E}\left[g^{(h)}(\underline{y}_t,\varepsilon_{t+1:t+h})H_K(\varepsilon_{t+1:t+h})\right]/\prod^h_{i=1}(k_i!).
\end{equation*}
Since $H_0(\varepsilon_{t+1:t+h})=1$ is a constant term, it has no effect on the conditional variance and the summation in \eqref{coro1} is taken over $\mathcal{K} = \left\{\mathbb{N}^4 \ \backslash \ \textbf{0}\right\}$ instead.  \qed\\

In the univariate dynamic framework, the HFEVD is a decomposition by time horizon and by degree of nonlinearity. Note that the coefficients $c^{(h)}_{K}(\underline{y_t})$ contain the superscript ${(h)}$ to indicate that they depend on forecast horizon $h$. They are also functions of the observed history $\underline{y_t}$; hence, the HFEVD is in general, not history invariant. For a better understanding of formula \eqref{fevd_quin_uni}, let us consider an autoregressive process at horizon 2: $Y_{t+2} = g^{(2)}(\underline{y_t},\varepsilon_{t+1},\varepsilon_{t+2})$. Then:
\begin{itemize}
	\item The term indexed by $K=(2,0)$ corresponds to the joint Hermite polynomial $H_K(\varepsilon_{t+1})=H_2(\varepsilon_{t+1})=\varepsilon_{t+1}^2-1$. 
	\item The term indexed by  $K=(0,2)$ corresponds to the joint Hermite polynomial $H_K(\varepsilon_{t+2})=H_2(\varepsilon_{t+2})=\varepsilon_{t+2}^2-1$. 
	\item The term indexed by $K=(1,1)$ corresponds to the joint Hermite polynomial  $H_K(\varepsilon_{t+1},\varepsilon_{t+2})=H_1(\varepsilon_{t+1})H_1(\varepsilon_{t+2})=\varepsilon_{t+1}\varepsilon_{t+2}$. 
\end{itemize}
Since the sum is taken over $\mathcal{K} = \left\{\mathbb{N}^h \ \backslash \ \textbf{0}\right\}$, the aggregation in \eqref{fevd_quin_uni} is potentially infinite, with the total number of terms in the expression depending solely on the degree of nonlinearity in $g^{(h)}$. A highly nonlinear trajectory will likely have many terms, while a more linear one could have most of the terms equal to 0\footnote{While this may seem discouraging for practitioners to implement, it is important to note that every term is orthogonal to one another and can be computed independently. In particular, the approximation accuracy of the entire trajectory does not influence the accuracy of individual terms within the approximation. Thus, a partial decomposition can still be obtained up to a finite number of terms without compromising the accuracy of the results.}. 

%When all higher order and joint Hermite polynomials have coefficients equal to 0, our HFEVD is equivalent to the classical linear FEVD. 

\subsection{Multivariate HFEVD}

The extension to the multivariate case follows naturally from the discussion above, but there are two key differences. First, the process $(Y_{t+h})$ is now of dimension $n$, and we now have a decomposition of the conditional variance-covariance matrix instead of a scalar conditional variance. Second, there are now $n\times h$ structural innovations across $h$ horizons. The proposition below extends the HFEVD to the nonlinear SVAR($p$) framework: 
\begin{proposition}[Multivariate HFEVD]
	Suppose the trajectory of an $n$-dimensional nonlinear SVAR($p$) is given by: 
	\begin{equation*}
		y_{t+h} = g^{(h)}(y_t,\varepsilon_{1,t+1:t+h},...,\varepsilon_{n,t+1:t+h}),
	\end{equation*}
	where each component of $g^{(h)}$ is square integrable and $\varepsilon_{j,t+1:t+h}=(\varepsilon_{j,t+1},...,\varepsilon_{j,t+h})$ for $j=1,..,n$ are independent standard Gaussian innovations. Its variance covariance matrix conditional on the observed history $\underline{y_t}$ is given by: 
%	\begin{equation}\label{fevd_quin_multi}
%		\mathbb{V}[Y_{t+h}|\underline{y_t}] \approx \sum_{k=1}^{\infty}\sum_{\substack{K\in\mathbb{N}^{nh}  \\ K:\textbf{1}'K=k}}\left[C^{(h)}_K(\underline{y_t})C^{(h)}_K(\underline{y_t})'\prod_{i=1}^{nh}(k_{i}!)^2\right],
%	\end{equation}
%where $C^{(h)}_K(\underline{y_t}) = \mathbb{E}\left[ g^{(h)}(y_t,\varepsilon_{1,t+1:t+h},...,\varepsilon_{n,t+1:t+h})H_{K}(\varepsilon_{1,t+1:t+h},...,\varepsilon_{n,t+1:t+h})\right]/\prod_{i=1}^{nh}(k_{i}!)$ and $ \textbf{1}'K=k_1+...+k_{nh}$.
	\begin{equation}\label{fevd_quin_multi}
	\mathbb{V}[Y_{t+h}|\underline{y_t}] \approx \sum_{K \in \mathcal{K}}\left[C^{(h)}_K(\underline{y_t})C^{(h)}_K(\underline{y_t})'\prod_{i=1}^{nh}(k_{i}!)^2\right],
\end{equation}
where $\mathcal{K} = \left\{\mathbb{N}^{nh} \ \backslash \ \textbf{0}\right\}$ and  $C^{(h)}_K(\underline{y_t}) = \mathbb{E}\left[ g^{(h)}(y_t,\varepsilon_{1,t+1:t+h},...,\varepsilon_{n,t+1:t+h})H_{K}(\varepsilon_{1,t+1:t+h},...,\varepsilon_{n,t+1:t+h})\right]/\prod_{i=1}^{nh}(k_{i}!)$. 
\end{proposition}

\noindent \textbf{Proof:} The proof is identical to Proposition 2, but now written in matrix form since we have a vector of coefficients. \qed \\

The vector $K=(k_{1,1},...,k_{n,h})$, which captures the degrees of nonlinearity for each marginal Hermite polynomial in the product, is now of dimension $n \times h$. The first $h$ elements correspond to the degrees of nonlinearity for $\varepsilon_{1,t+1:t+h}$, the next $h$ elements correspond to the degrees of nonlinearity for $\varepsilon_{2,t+1:t+h}$, ..., and the last $h$ elements correspond to the degrees of nonlinearity for $\varepsilon_{n,t+1:t+h}$. Moreover, the HFEVD is now a decomposition of the conditional variance-covariance matrix along three dimensions: by time horizon, by structural innovation components and by degree of nonlinearity. To illustrate this point, let us consider a bivariate autoregressive process at horizon $h=2$. Then:
\begin{itemize}
	\item The term indexed by $K=(2,0,0,0)$ corresponds to the Hermite polynomial $H_K(\varepsilon_{t+1:t+h})=H_2(\varepsilon_{1,t+1})=\varepsilon_{1,t+1}^2-1$. 
	\item The term indexed by $K=(0,0,0,2)$ corresponds to the Hermite polynomial $H_K(\varepsilon_{t+1:t+h})=H_2(\varepsilon_{2,t+2})=\varepsilon_{2,t+2}^2-1$. 
	\item The term indexed by $K=(1,1,0,0)$ corresponds to the Hermite polynomial $H_K(\varepsilon_{t+1:t+h})=H_1(\varepsilon_{1,t+1})H_1(\varepsilon_{1,t+2})=\varepsilon_{1,t+1}\varepsilon_{1,t+2}$. 
	\item The term indexed by $K=(0,0,1,1)$ corresponds to the Hermite polynomial $H_K(\varepsilon_{t+1:t+h})=H_1(\varepsilon_{2,t+1})H_1(\varepsilon_{2,t+2})=\varepsilon_{2,t+1}\varepsilon_{2,t+2}$. 
\end{itemize}
%In total, there will be $\binom{5}{3}=10$ such terms for $k=2$ and the number of terms for any arbitrary $k \geq 1$ will be $\binom{k+3}{3}$ \footnote{More precisely, at horizon $h$ there will be $n \times h$ and we have to distribute $k$ degrees among them. Hence, we have $n \times h$ ``bins" and $k$ ``balls", and the generalized formula is given by: $\binom{k+nh-1}{nh-1}$}.

\subsection{Interpretation of the HFEVD}

In practice, appropriate terms in the Hermite polynomial expansion can be selected in the decomposition to obtain an interpretation that is of interest to the practitioner. For exposition in this section, consider the bivariate nonlinear autoregressive process $Y_{t}=(Y_{1,t},Y_{2,t})'$. At horizon 2, its trajectory is given by:
\begin{equation}\label{exh2}
	Y_{t+2} = g^{(2)}(\underline{y_t},\varepsilon_{1,t+1},\varepsilon_{1,t+2},\varepsilon_{2,t+1},\varepsilon_{2,t+2}).
\end{equation}
By applying \eqref{fevd_quin_multi}, the conditional variance of \eqref{exh2} is of the form: 
\begin{equation}\label{vexp}
	\mathbb{V}\left[Y_{t+2}|\underline{y_t}\right]= \sum_{K \in \mathcal{K}}\mathbb{VC}^{(2)}_K(y_t) =\sum_{k_{11}} \sum_{k_{12}} \sum_{k_{21}}  \sum_{k_{22}}\mathbb{VC}^{(2)}_{(k_{11},k_{12},k_{21},k_{22})}(y_t),
\end{equation}
where $\mathbb{VC}^{(2)}_K(y_t)=\left[C_K^{(2)}(\underline{y_t})C_K^{(2)}(\underline{y_t})'(k_{11}!k_{12}!k_{21}!k_{22}!)^2\right]$ denotes the variance contribution of the joint Hermite polynomial indexed by $K=(k_{11},k_{12},k_{21},k_{22})$, $K\neq \textbf{0}$, and $k_{11}$, $k_{12}$, $k_{21}$ and $k_{22}$ represent the degrees of nonlinearity for the marginal Hermite polynomials of the innovations $\varepsilon_{1,t+1}$, $\varepsilon_{1,t+2}$, $\varepsilon_{2,t+1}$ and $\varepsilon_{2,t+2}$, respectively.

\subsubsection{Types of Contributions}

 Each term in \eqref{vexp} can be classified as a contribution to the conditional variance from either specific innovations or interactions between multiple innovations. For instance, the $k_{11}$-th degree variance contribution specific to innovation $\varepsilon_{1,t+1}$ is given by:  
\begin{equation}\label{vc1}
	\mathbb{VC}^{(2)}_{(k_{11},0,0,0)}(y_t) = \left[C_{(k_{11},0,0,0)}^{(2)}(\underline{y_t})C_{(k_{11},0,0,0)}^{(2)}(\underline{y_t})'(k_{11})!^2\right].
\end{equation}
When $k_{11}=1$, \eqref{vc1} captures the linear contribution specific to innovation $\varepsilon_{1,t+1}$. Likewise, $k_{11}\geq 2$ captures specific higher-order contributions of $\varepsilon_{1,t+1}$, such as quadratic (for $k_{11}=2$) or cubic (for $k_{11}=3$). Contributions can also be defined jointly for multiple innovations and arise due to interactions between them. For instance, the joint contribution between $\varepsilon_{1,t+1}$ and $\varepsilon_{2,t+1}$ at degrees of nonlinearity $k_{11}$ and $k_{21}$, respectively, is given by: 
\begin{equation}
	\mathbb{VC}^{(2)}_{(k_{11},0,k_{21},0)}(y_t) = \left[C_{(k_{11},0,k_{21},0)}^{(2)}(\underline{y_t})C_{(k_{11},0,k_{21},0)}^{(2)}(\underline{y_t})'(k_{11}!k_{21}!)^2\right],
\end{equation}
for $k_{11},k_{21} \geq 1$. These contributions can also be aggregated in different ways. 
\begin{enumerate}
	\item \textbf{Linear Contributions:} A practitioner may want to assess whether shock(s) in their model exhibit linear behaviour. This can be done by computing the total linear contributions of the innovations:
	\begin{equation}\label{linc}
		\mathbb{VC}_{(1,0,0,0)}^{(2)}(\underline{y_t})+	\mathbb{VC}_{(0,1,0,0)}^{(2)}(\underline{y_t})+	\mathbb{VC}_{(0,0,1,0)}^{(2)}(\underline{y_t})+	\mathbb{VC}_{(0,0,0,1)}^{(2)}(\underline{y_t}).
	\end{equation}
	\item \textbf{Marginal Contributions:} The marginal contribution of a specific innovation is the sum of all its linear and higher-order effects. For instance, the marginal contribution of $\varepsilon_{1,t+1}$ is given by:
	\begin{equation}\label{mc2}
		\sum_{k_{11}=1}^\infty \mathbb{VC}^{(2)}_{(k_{11},0,0,0)}(\underline{y_t}), 
	\end{equation}  
	which aggregates the linear, quadratic, cubic, ..., contributions of $\varepsilon_{1,t+1}$. 
	\item \textbf{Joint Contributions:} The joint contributions of an innovation is the sum of all its interactions with other innovations. For instance, the joint contributions of $\varepsilon_{1,t+1}$ can be written as:
		\begin{equation}\label{mc}\footnotesize
\begin{split}
		&	\underbrace{\sum_{k_{11}=1}^\infty\sum_{k_{12}=1}^\infty \mathbb{VC}^{(2)}_{(k_{11},k_{12},0,0)}(\underline{y_t})+ \sum_{k_{11}=1}^\infty\sum_{k_{21}=1}^\infty \mathbb{VC}^{(2)}_{(k_{11},0,k_{13},0)}(\underline{y_t})+ \sum_{k_{11}=1}^\infty\sum_{k_{22}=1}^\infty \mathbb{VC}^{(2)}_{(k_{11},0,0,k_{14})}(\underline{y_t})}_{\text{Pairwise Interactions}}\\
	+	&\underbrace{ \sum_{k_{11}=1}^\infty\sum_{k_{12}=1}^\infty\sum_{k_{12}=1}^\infty \mathbb{VC}^{(2)}_{(k_{11},k_{12},k_{21},0)}(\underline{y_t})+ \sum_{k_{11}=1}^\infty\sum_{k_{12}=1}^\infty\sum_{k_{22}=1}^\infty \mathbb{VC}^{(2)}_{(k_{11},k_{12},0,k_{22})}(\underline{y_t})+ \sum_{k_{11}=1}^\infty\sum_{k_{21}=1}^\infty\sum_{k_{22}=1}^\infty \mathbb{VC}^{(2)}_{(k_{11},0,k_{21},k_{22})}(\underline{y_t})}_{\text{Triplet-wise Interactions}}\\
	+ & \underbrace{\sum_{k_{11}=1}^\infty\sum_{k_{12}=1}^\infty\sum_{k_{21}=1}^\infty\sum_{k_{22}=1}^\infty \mathbb{VC}^{(2)}_{(k_{11},k_{12},k_{21},k_{22})}(\underline{y_t})}_{\text{Quadruplet-Wise Interactions}}.\\
\end{split}
	\end{equation}  
\end{enumerate}

\subsubsection{Interpretable Partitions}

The set $\mathcal{K}=\mathbb{N}^4 \ \backslash \ \left\{\textbf{0}\right\}$, which contains all possible indices in the expansion for \eqref{ex_1}, can be partitioned in different ways to obtain interpretations of the HFEVD that are of interest to the practitioner. Moreover, since each term in the decomposition are orthogonal to one another, these partitions can be easily computed even if the trajectory has many high-order nonlinear terms. For instance, suppose $\mathcal{K} = \mathcal{K}_1 \bigoplus \mathcal{K}_2$ can be partitioned into two sets and consider the following partitions. 
\begin{itemize}
	\item Let $\mathcal{K}_1 =\left\{(1,0,0,0),(0,1,0,0),(0,0,1,0),(0,0,0,1)\right\}$ and $\mathcal{K}_2 = \mathcal{K} \ \backslash \ \mathcal{K}_1$. Then, the sum of terms indexed by $\mathcal{K}_1$ are the linear contributions of all innovations in the model as computed in \eqref{vc1}, and the terms indexed by $\mathcal{K}_2$ will consist of the contributions from all higher-order and interaction terms.
	\item Let $\mathcal{K}_1 =\left\{K \in \mathcal{K} \ \vert \ k_{11}\geq 1 \right\} \cup \left\{K \in \mathcal{K} \ \vert \ k_{12}\geq 1 \right\}$ and $\mathcal{K}_2 = \mathcal{K} \ \backslash \ \mathcal{K}_1$. Then, the sum of terms indexed by $\mathcal{K}_1$ consists of all Hermite polynomials of the innovation components, i.e. $\varepsilon_{1,t+1}$ or $\varepsilon_{1,t+2}$, including their marginal and joint contributions. Note that this partition coincides with the nonlinear FEVD:
	\begin{equation}
	\mathbb{V}[Y_{t+2}|\underline{y_t}]=	\mathbb{E}[\mathbb{V}(Y_{t+2}|\varepsilon_{2,t+1},\varepsilon_{t+2},\underline{y_t})]+\mathbb{V}[\mathbb{E}(Y_{t+2}|\varepsilon_{2,t+1},\varepsilon_{t+2}, \underline{y_t})],
	\end{equation}
	as proposed in \cite{IN2020}. They argue that  $\mathbb{E}[\mathbb{V}(Y_{t+2}|\varepsilon_{2,t+1},\varepsilon_{t+2},\underline{y_t})]$ captures not only the ``direct" (marginal) effects  of a shock on the first component, but also its ``indirect" (interaction) effects. In the context of the HFEVD, this is equivalent to including the marginal and joint contributions of an innovation [To see this explicitly in a numerical context, see the example presented in Section 4.6 below]\footnote{A different partitioning can provide a more granular separation than the FEVD of \cite{IN2020}. For instance, suppose $\mathcal{K} = \mathcal{K}_1 \bigoplus \mathcal{K}_2 \bigoplus \mathcal{K}_3$ is now partitioned into three subsets.  Let $\mathcal{K}_1 =\left\{K \in \mathcal{K} \ \vert \ k_{11}\geq 1, k_{12}=k_{21}=k_{22}=0\right\} \cup \left\{K \in \mathcal{K} \ \vert \ k_{12}\geq 1, k_{11}=k_{21}=k_{22}=0  \right\}$, which captures the marginal contributions of the first innovation component. Let $\mathcal{K}_2 =\left\{K \in \mathcal{K} \ \vert \ k_{11}\geq 1 \right\} \cup \left\{K \in \mathcal{K} \ \vert \ k_{12}\geq 1 \right\} \ \backslash \ \mathcal{K}_1$. This captures the joint contributions of the first innovation component. Finally, let $\mathcal{K}_3 = \mathcal{K} \ \backslash \ \left\{\mathcal{K}_1 \ \mathcal{K}_2\right\}$. Then, the ``direct" and ``indirect" effects are distinguished by the terms indexed by $\mathcal{K}_1$ and $\mathcal{K}_2$. }. 	
\end{itemize}

\subsection{Nonlinear EIRF and HFEVD}

In Section 2.3, a relationship between IRFs and FEVDs was established. An analogous relationship exists between nonlinear EIRFs and the HFEVD. Akin to the linear framework, we can define a perturbed path $Y_{t+h}(\delta)$ and a baseline path $Y_{t+h}$ as: 
\begin{equation}\label{nonlinear_irf_paths}
	\begin{split}
		Y_{t+h} &= g^{(h)}(\underline{y_{t}},\varepsilon_{t+1:t+h}), \ \forall h, \\
		Y_{t+h}(\delta) &= g^{(h)}(\underline{y_{t}},\varepsilon_{t+1}+\delta,\varepsilon_{t+2:t+h}),  \ \forall h.\\
	\end{split}
\end{equation}
The uncertainty of the nonlinear Impulse Response Function (IRF) can be summarized by the joint distribution of these two paths [\cite{GJ2005}, Definition 4]. In the current literature, it is standard practice to consider the expected difference of these paths, conditional on a history $\underline{y_{t}}$:
\begin{equation}\label{EIRF}
	EIRF(h,\delta,\underline{y_t}) = \mathbb{E}[Y_{t+h}(\delta) - 	Y_{t+h}|\underline{y_t}]. 
\end{equation}
For exposition, let us consider the EIRF of a univariate process at horizon 1. Then $Y_{t+1}=g(\underline{y_t},\varepsilon_{t+1})$ will be a function of a single Gaussian innovation $\varepsilon_{t+1}$. A Taylor expansion of the EIRF yields the following: 
\begin{lemma}
\begin{equation}\label{eirf2}
EIRF(1,\delta,\underline{y_t})=\mathbb{E}[Y_{t+1}(\delta) - 	Y_{t+1}|\underline{y_t}] = \sum_{k=1}^\infty \frac{\delta^k}{k!}\mathbb{E}\left[\frac{d^k}{d\varepsilon_{t+1}^k}g(\underline{y_t},\varepsilon_{t+1})\right].
\end{equation}
\end{lemma} 

\noindent \textbf{Proof:} See Appendix A.1. \\

Lemma 3 provides an intepretation of the EIRF as implied by the perturbed and baseline paths in \eqref{nonlinear_irf_paths}. Each term in the summation is the product of the $k$-th order Gaussian marginal effect of $\varepsilon_{t+1}$ on $Y_{t+h}$ and a scaling of the shock magnitude.  In the linear SVAR model discussed in \eqref{SVAR}, $EIRF(1,\delta,\underline{y_t})=\delta AD$, and the marginal effect of $\varepsilon_{t+1}$ on $Y_{t+h}$ is the impact multiplier $\Theta_i=AD=\frac{dY_{t+h}}{d\varepsilon_{t+1}}$. Hence, we can interpret \eqref{eirf2} as a weighted sum of impact multipliers at all degrees of nonlinearity $k$. Morever, these marginal effects are linked directly to the terms in the HFEVD in the following way\footnote{Two alternative defintions of the IRF have also been considered in the past: [1] The ``MIT" shock where a shock is characterized by $\varepsilon_{t+1}+\delta$ as in this paper, but all other innovations are set directly equal to 0. [2] The Generalized Impulse Response Function (GIRF), where a shock is characterized by setting the innovation directly equal to the magnitude of the shock, i.e. $\varepsilon_{t+1}=\delta$ [\cite{KPP1996}]. Their approaches do not share the same interpretation as Lemma 3 and are not appropriate characterizations of shocks in nonlinear dynamic models [see Appendix B for details].}:

\begin{proposition}
The HFEVD of a univariate process at horizon $1$ is given by:
	\begin{equation}
	\mathbb{V}[Y_{t+1}|\underline{y_t}] = \sum_{k=1}^{\infty}\left[c^{(h)}_k(\underline{y_t}) (k!)\right]^2,
\end{equation}
where:
\begin{equation}
\left[c^{(h)}_k(\underline{y_t}) (k!)\right]^2= \mathbb{E}\left[\frac{d^k}{d\varepsilon_{t+1}^k}g(\underline{y_t},\varepsilon_{t+1})\right]^2,
\end{equation}
for all $k \geq 1$.
\end{proposition}

\noindent \textbf{Proof:} See Appendix A.2. \\

By analogy to the relationship in the linear framework, Proposition 4 shows that the terms in the HFEVD are equivalent to the squared impact multipliers of the IRF expansion in \eqref{eirf2}. This result can be extended to multivariate processes as well. Without loss of generality, consider the case of a bivariate process at horizon 1. Then $Y_{t+1}= g(\underline{y_t},\varepsilon_{1,t+1},\varepsilon_{2,t+1})$ is a function of Gaussian innovations $\varepsilon_{1,t+1}$ and $\varepsilon_{2,t+2}$ and we get the following proposition: 
\begin{proposition}
The HFEVD of a bivariate process at horizon $1$ is given by: 
\begin{equation}
	\mathbb{V}[Y_{t+h}|\underline{y_t}] =\sum_{K \in \mathcal{K}_2}\left[C^{(h)}_K(\underline{y_t})C^{(h)}_K(\underline{y_t})'(k_1!k_2!)^2\right]
\end{equation}
where $\mathcal{K}_2=\mathbb{N}^2 \ \backslash \ \textbf{0}$ and: 
\begin{equation}
	\begin{split}
		C_K(\underline{y_t})(k_1!k_2!)= \mathbb{E}\left[\frac{\partial^{(k_1+k_2)}}{\partial\varepsilon_{1,t+1}^{k_1}\partial \varepsilon_{2,t+1}^{k_2}}g(\underline{y_t},\varepsilon_{1,t+1},\varepsilon_{2,t+1})\right].
	\end{split}
\end{equation}
\end{proposition}

\noindent \textbf{Proof:} See Appendix A.3. \\ 

The terms in the EIRF expansion now feature cross-derivatives between $\varepsilon_{1,t+1}$ and $\varepsilon_{2,t+1}$. They capture the interaction effects between shocks, and are linked to the joint Hermite polynomials between the structural innovations. Hence, we can interpret the contribution of marginal Hermite polynomials as the expected marginal effects of shocks, and the contribution of joint Hermite polynomials as their interaction effects. These results can be generalized to any dimension $n$ and horizon $h$, where the expansion of the EIRF will contain cross-derivatives of the $n\times h$ innovations. \\

\noindent \textbf{Remark 4:} As shown in Appendix D.4, the link between the nonlinear FEVD and EIRF only holds for Gaussian innovations. Hence, Gaussianity is a very special case of economic innovations, and non-Gaussianity is in fact a source of nonlinearity that differs from the nonlinearity introduced by the dynamics of a model (e.g. An AR(1) process with t(5) errors is linear in the modelling dynamic, but exhibits nonlinear behaviour due to the non-Gaussianity of the innovation). Moreover, this observation also tells us that any nonlinear FEVD constructed based on the definition of the nonlinear IRF, such as the GFEVD approach of \cite{LN2016}, can have misleading interpretations when errors are non-Gaussian. 

\subsection{Example}

Consider an explicit calculation of the HFEVD for the nonlinear process given by:
\begin{equation*}
	\begin{split}
		y_{1,t} & = a y_{1,t-1} + y_{2,t-1}\varepsilon^2_{1,t},\\
		y_{2,t} & = by_{2,t-1} + \varepsilon_{2,t},\\
	\end{split}
\end{equation*}
where $\varepsilon_t$ is a bivariate standard Gaussian noise. At horizon $h$ the value of this process is:
\begin{equation*}
	\begin{split}
		y_{1,t+h} & = a^hy_{1,t} + a^{h-1}y_{2,t}\varepsilon_{1,t+1} + \sum_{k=2}^{h-k}\left(b^ky_{2,t}+\sum_{k}^{k}b^{}\varepsilon_{2,t+1}\right)^2\varepsilon_{1,t+k},\\
		y_{2,t+h} & = b^hy_{2,t} + \sum_{k=1}^h b^{h-k} \varepsilon_{2,t}. \\
	\end{split}
\end{equation*}
Suppose we are interested in the decomposition at horizon $t+3$. Then: 
\begin{equation}\label{ex_1}
	\begin{split}
		y_{1,t+3} & = a^3y_{1,t}+a^2y_{2,t}\varepsilon^2_{1,t+1}+a(by_{2,t}+\varepsilon_{2,t+1})\varepsilon^2_{1,t+2}+(b^2y_{2,t+1}+b\varepsilon_{2,t+1}+\varepsilon_{2,t+2})\varepsilon^{2}_{1,t+3},\\
		y_{2,t+3} & = b^3y_{2,t}+b^2\varepsilon_{2,t+1}+b\varepsilon_{2,t+2}+\varepsilon_{2,t+3}. \\
	\end{split}
\end{equation}
There are 3 (horizons) $\times$ 2 (variables) innovations in the system. For $Y_{1,t+3}$, there will be at most quadratic effects of $\varepsilon_{1,t+l}$ and linear effects of $\varepsilon_{2,t+l}$. For $y_{2,t+3}$, it is a function of only linear terms of $\varepsilon_{2,t+l}$. Thus, it suffices to consider Hermite polynomials of degree 1 and 2 for this decomposition that only include the marginal and pairwise contributions in the total variance. The resulting expansion is given by: 
\begin{equation}\label{ql_ex_1}
	\begin{split}
		Y_{t+3} &= \begin{bmatrix}
			a^3y_{1,t}+a^2y_{2,t}+aby_{2,t}+b^2y_{2,t} \\b^3y_{2,t}\\ 
		\end{bmatrix}\times 1 +
		\begin{bmatrix}
			a^2y_{2,t} \\ 0 \\ 
		\end{bmatrix}\times (\varepsilon^2_{1,t+1}-1)+
		\begin{bmatrix}
			aby_{2,t} \\ 0\\ 
		\end{bmatrix}\times (\varepsilon^2_{1,t+2}-1) \\ 
		& + \begin{bmatrix}
			b^2y_{2,t} \\ 0 \\ 
		\end{bmatrix}\times (\varepsilon^2_{1,t+3}-1) 
		+ \begin{bmatrix}
			a+b \\ b^2 \\ 
		\end{bmatrix}\times (\varepsilon_{2,t+1}) 
		+ \begin{bmatrix}
			1\\ b\\ 
		\end{bmatrix}\times (\varepsilon_{2,t+1}) 
		+ \begin{bmatrix}
			0\\ 1 \\ 
		\end{bmatrix}\times (\varepsilon_{2,t+1}) \\ 
		& + \begin{bmatrix}
			a \\ 0 \\ 
		\end{bmatrix}\times (\varepsilon^2_{1,t+2}-1)\varepsilon_{2,t+1}
		+ \begin{bmatrix}
			b \\ 0 \\ 
		\end{bmatrix}\times (\varepsilon^2_{1,t+3}-1)\varepsilon_{2,t+1}
		+ \begin{bmatrix}
			1 \\ 0 \\ 
		\end{bmatrix}\times (\varepsilon^2_{1,t+3}-1)\varepsilon_{2,t+2}
	\end{split}
\end{equation}
which is  equivalent to the partial nonlinear moving representation \eqref{ex_1}. This expression separates the effects of shocks on $Y_{t+3}$ by time horizon, by structural innovation component and by degree of nonlinearity. For example, the term $(\varepsilon^2_{t+3}-1)$ captures the quadratic effect of innovation $\varepsilon_1$ at horizon $3$. Its influence on $y_{1,t+3}$ is $b^2y_{2,t}$, but it has no effect on $y_{2,t+3}$. We also see interaction effects between innovations, for instance, in the term $(\varepsilon^2_{1,t+3}-1)\varepsilon_{2,t+2}$. 
Since the expansion above is now a sum of orthogonal Hermite polynomials, the conditional variance-covariance matrix of $Y_{t+3}$ can be additively separated as the sum of the individual variance-covariance matrices corresponding to the difference terms. It can be shown that:
\begin{equation}\label{ql_ex_1_var}
	\begin{split}
		\mathbb{V}[Y_{t+3}|\underline{y_t}] & = 
		\underbrace{\begin{bmatrix}
				a^4y^2_{2,t} & 0 \\0 & 0 \\ 
			\end{bmatrix}\times 2}_{\text{Quadratic Contribution}, \varepsilon_{1,t+1}}+
		\underbrace{\begin{bmatrix}
				a^2b^2y^2_{2,t} & 0 \\0 & 0\\ 
			\end{bmatrix}\times 2}_{\text{Quadratic Contribution}, \varepsilon_{1,t+2}} 
		+ \underbrace{\begin{bmatrix}
				b^4y^2_{2,t} & 0\\0& 0 \\ 
			\end{bmatrix}\times  2 }_{\text{Quadratic Contribution}, \varepsilon_{1,t+3}}\\
		& 
		+\underbrace{ \begin{bmatrix}
				(a+b)^2 & (a+b)b^2 \\b^2(a+b) & b^4 \\ 
		\end{bmatrix} }_{\text{Linear Contribution}, \varepsilon_{2,t+1}}
		+ \underbrace{\begin{bmatrix}
				1 & b\\ b &  b^2\\ 
		\end{bmatrix}}_{\text{Linear Contribution}, \varepsilon_{2,t+2}}
		+ \underbrace{\begin{bmatrix}
				0 & 0 \\ 0 &  1 \\ 
		\end{bmatrix} }_{\text{Linear Contribution}, \varepsilon_{2,t+3}} \\
		& +\underbrace{ \begin{bmatrix}
				a^2 & 0 \\0 & 0 \\ 
			\end{bmatrix}\times 2}_{\substack{\text{Linear Contribution}, \varepsilon_{2,t+1} \\ \times  \text{Quadratic Contribution}, \varepsilon_{1,t+2} }}
		+ \underbrace{\begin{bmatrix}
				b^2 & 0  \\ 0& 0 \\ 
			\end{bmatrix}\times 2}_{\substack{\text{Linear Effect},\varepsilon_{2,t+1} \\ \times  \text{Quadratic Contribution}, \varepsilon_{1,t+3} }}
		+\underbrace{ \begin{bmatrix}
				1 & 0 \\ 0 & 0 \\ 
			\end{bmatrix}\times 2}_{\substack{\text{Linear Contribution}, \varepsilon_{2,t+2} \\ \times \text{Quadratic Contribution}, \varepsilon_{1,t+3} }}
	\end{split}
\end{equation}
The diagonal elements of each matrix correspond to the contribution of each innovation to the variance of $y_{1,t+3}$ and $y_{2,t+3}$. For example, the linear effects of all $\varepsilon_2$ innovations on $y_{1,t+3}$ is given by $(a+b)^2+1$ (the fourth and fifth term of the FEVD). On the other hand, there are no linear effects from the $\varepsilon_1$ innovations. The non-diagonal elements represent the contribution of each innovation to the covariance between $y_{1,t+3}$ and $y_{2,t+3}$. For example, any term with $\varepsilon_1$ (whether joint or marginal) has no effect, since $\varepsilon_1$ only appears in the equation for $y_{1,t+3}$ but not $y_{2,t+2}$. We also note that the covariance contributions can be negative and depends on the parameters $(a,b)$. If both $a$ and $b$ are both negative, then the linear effects of $\varepsilon_2$ on the covariance is also negative, since $b^2(a+b)+b <0$.\\

\noindent \textbf{Remark 5:} \cite{IN2020} argue that  $\mathbb{E}_t[\mathbb{V}(y_{1,t+h}|\varepsilon^{-1}_{t+1:t+3})]=\mathbb{E}_t[\mathbb{V}(y_{1,t+h}|\varepsilon_{2,t+1},\varepsilon_{2,t+2},\varepsilon_{2,t+3})]$ captures not only the marginal contribution of innovation 1, but the contribution of its interactions. Indeed:
\begin{equation}\label{in_ex_1}
	\begin{split}
		&\mathbb{E}_t[\mathbb{V}(y_{1,t+h}|\varepsilon_{2,t+1},\varepsilon_{2,t+2},\varepsilon_{2,t+3})]\\
		= &\underbrace{2a^4 y^2_{2,t}}_{\text{Quadratic Contribution}, \varepsilon_{1,t+1}} + \underbrace{2a^2b^2y_{2,t}^2}_{\text{Quadratic Contribution}, \varepsilon_{1,t+1}}+\underbrace{2a^2}_{\substack{\text{Linear Contribution}, \varepsilon_{2,t+1} \\ \times  \text{Quadratic Effect}, \varepsilon_{1,t+2} }}\\ 
		+ & \underbrace{2b^4y_{2,t}^2}_{\text{Quadratic Contribution}, \varepsilon_{1,t+3}} + \underbrace{2b^2}_{\substack{\text{Linear Effect}, \varepsilon_{2,t+1} \\ \times  \text{Quadratic Contribution}, \varepsilon_{1,t+3} }} + \underbrace{2}_{\substack{\text{Linear Contribution}, \varepsilon_{2,t+2} \\ \times  \text{Quadratic Contribution}, \varepsilon_{1,t+3} }}\\
	\end{split}
\end{equation}
These terms coincide with the marginal and joint Hermite polynomials of $\varepsilon_{1,t+k}$ for $k=1,2,3$. Hence, the Isakin and Ngo (2020) approach is a special case of the HFEVD. 

\section{Numerical Implementation}

This section focuses on how the HFEVD can be implemented in practice. There are two issues that may arise with the proposed approach. First, the future trajectory in \eqref{pvolt} may not be available in closed form. Second, the model may be semi-parametric in the sense that the i.i.d. innovations are not necessarily Gaussian. The parametric framework is discussed first followed by the semi-parametric framework. In both cases, it is assumed that the parameter $\theta$ is identifiable. 

\subsection{Parametric Framework}

Consider the $n$ dimensional nonlinear SVAR(p) process described in Proposition 1:
\begin{equation*}
	Y_t = g(\underline{Y_{t-1}},\varepsilon_t;\theta),
\end{equation*}
where $\varepsilon_t$ is assumed to be standard Gaussian and $\theta$ denotes a parameter which governs the function $g$. To evaluate some of the contributions to the conditional total variance, the approach follows the steps below:

\begin{enumerate}
	\item Estimate $\theta$ by maximum likelihood from the observations $y_1,...,y_T$\footnote{The estimator would have to be indexed by $T$ to mention its dependence on observation, but we omit this for expository purposes.}. Then, deduce $\hat{g}_T(\underline{Y_{t-1}},\varepsilon_t)=g(\underline{Y_{t-1}},\varepsilon_t,\hat{\theta}_T)$. \footnote{The HFEVD obtained in this numerical implementation depends on the estimate of function $g$ (or parameter $\theta$), so it is also subject to estimation risk, which is discussed in Appendix C.}.
	\item Simulate $S$ draws from a standard Gaussian distribution to obtain $\varepsilon^s_{t+i}$ for $s=1,...,S$ and $i=1,...,h$. 
	\item Simulate the future trajectory $S$ times, where each simulation is given by:
	\begin{equation}
		y_{t+h}^s = \hat{g}_T^{(h)}(\underline{y_t},\varepsilon^s_{t+1},...,\varepsilon^s_{t+h})
	\end{equation} The value $\underline{y_t}$ is chosen either based on the practitioner's interest on a particular historical profile, or by taking the last observations available in the data set. 
	\item Construct the Hermite polynomials $H_K(\varepsilon^s_{t+1},...,\varepsilon^s_{t+h})$ of the simulated Gaussian innovations for all $K=(k_1,...,k_{nh})$ in the set $\Omega \subset \mathbb{N}^{nh}$ of interest.
		\item Compute the (estimated) Hermite coefficients:
	\begin{equation}
		\tilde{C}_K(\underline{y_t}) = \frac{1}{S}\sum_{s=1}^S\left[y_{t+h}^sH_K(\varepsilon^s_{t+1},...,\varepsilon^s_{t+h})]/\prod_{i=1}^{nh}(k_{i}!)\right].
	\end{equation}
	\item Compute the variance contribution for the terms of interest:
	\begin{equation}
		\sum_{\substack{K\in \Omega}}\left[\tilde{C}_K(\underline{y_t})\tilde{C}_K(\underline{y_t})'\prod_{i=1}^{nh}(k_{i}!)^2\right].
	\end{equation}
\end{enumerate}

\subsection{Semiparametric Model}

Consider the $n$ dimensional SVAR(p) process described in \eqref{NLAR_alt}:
\begin{equation*}
	y_t = g^*(\underline{y_{t-1}},u_t;\theta) \iff u_t(\theta) = g^{*-1}(\underline{y_t};\theta).
\end{equation*}
We assume that this model is well-specified with a true value $\theta_0$ of the parameter, and that $u_t = u_t(\theta_0)$ is i.i.d. with a distribution $F_0$ which is unknown to the practitioner and not necessarily Gaussian. 

\begin{enumerate}
	\item Estimate $\theta$ by minimizing a covariance distance, for instance, by using the Generalized Covariance (GCov) approach [\cite{GJ2023}] written on tranformations of $u_t(\theta)$ and $u_{t-1}(\theta)$. If the model is linear in i.i.d. errors $u_t$, then it is also possible to apply a Psuedo Maximum Likelihood approach [\cite{GMR2017}]. Then, compute the residuals: $\hat{u}_{T,t} = g^{*-1}(\underline{y_t},\hat{\theta}_T)$. 
 	\item Simulate (bootstrap) the future trajectory $S$ times, where each simulation is given by:
 	\begin{equation}
 		y_{t+h}^s = \hat{g}_T^{(h)}(\underline{y_t},u^s_{t+1},...,u^s_{t+h}),
	 \end{equation} 
 and the $u^s_{t+i}$ are independently drawn from the sample distribution of $\hat{u}_{T,t}$, $t=1,...,T$.
 The value $\underline{y_t}$ is chosen either based on the practitioner's interest, on a particular historical profile, or by taking the last observations available in the data set. 
	\item Construct the simulated Gaussian innovations:
	\begin{equation}
		\varepsilon^s_{t+i} = \Phi^{-1}\hat{F}_T(u^s_{t+i}), 
	\end{equation}
	where $\hat{F}_T$ is the sample c.d.f. of $\hat{u}_{T,t}$, $t=1,...,T$\footnote{In practice for instance, it can be estimated by the empirical C.D.F. $\hat{F}_T(x)=\frac{1}{T}\sum_{i=1}^T \textbf{1}_{X_i \leq x}$.}
	\item Construct the Hermite polynomials $H_K(\varepsilon^s_{t+1},...,\varepsilon^s_{t+h})$ of the simulated Gaussian innovations for all $K=(k_1,...,k_{nh})$ in the set $\Omega \subset \mathbb{N}^{nh}$ of interest.
	\item Compute the coefficients:
	\begin{equation}
		\tilde{C}_K(\underline{y_t}) = \frac{1}{S}\sum_{s=1}^S\left[y_{t+h}^sH_K(\varepsilon^s_{t+1},...,\varepsilon^s_{t+h})]/\prod_{i=1}^{nh}(k_{i}!)\right].
	\end{equation}
	\item Compute the variance contribution for the terms of interest:
	\begin{equation}
		\sum_{\substack{K\in \Omega}}\left[\tilde{C}_K(\underline{y_t})\tilde{C}_K(\underline{y_t})'\prod_{i=1}^{nh}(k_{i}!)^2\right].
	\end{equation}
\end{enumerate}

\section{Simulation Exercises}

In this section, we provide the HFEVD analysis of simulated data following the examples discussed in Section 3.3.

\subsection{Double Autoregressive Model}

Consider the Double Autoregressive model in Section 3.3.1 with parameter values $\alpha =1$, $\phi = 0.5$ and $\beta=0.5$, which ensures a strictly stationary solution with finite second-order moments \footnote{Ling (2007) describes the stationarity region as the set $\{(\phi,a) \ s.t. \ \mathbb{E}|\ln\phi+\sqrt{a}\varepsilon_t|<0$\}.}. This yields the data generating process: 
\begin{equation*}
	y_t = 0.5 y_{t-1} + \sqrt{1 +0.5 y_{t-1}^2} u_t.
\end{equation*}
For the structural innovations we consider the cases $u_t \sim N(0,1)$ and $u_t \sim t(5)$. Then, the algorithms for the parameteric model (Section 5.1) and semi-parametric model (Section 5.2) can be applied to these two cases respectively. First, simulations of 200 observations for each process with initial value $y_0 = 0$ are presented in the two graphs of Figure 1. As expected, the variation for the DAR(1) model with Gaussian innovations exhibit much lower variability than the DAR(1) model with t(5) innovations.

\begin{figure}[h!]
	\centering
	\includegraphics[width=0.85\linewidth]{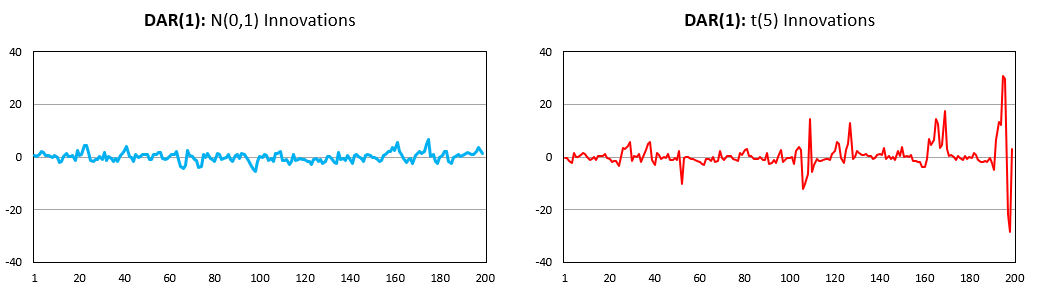}
	\caption{Simulated data (T=200) of DAR(1) processes with N(0,1) innovations (left graph) and t(5) innovations (right graph). }
	\label{fig:darfigure1}
\end{figure}

Figure 2 examines the linear contributions of innovations on the conditional total variance of the DAR(1) at the conditioning value $Y_T=0$. The blue and red bars represent the results for processes with N(0,1) and (normalized) t(5) structural innovations, respectively. As the forecast horizon increases, the linear contributions decrease more rapidly for the processes with t(5) innovations. This difference arises from the presence of fatter tails in the t-distribution, which introduces a nonlinear effect even though the underlying dynamics of the processes remain the same. Consequently, if the data were incorrectly modeled using a linear AR(1) process, it could lead to significant misinterpretations of the FEVD and potentially erroneous conclusions.

\begin{figure}[h!]
	\centering
	\includegraphics[width=0.65\linewidth]{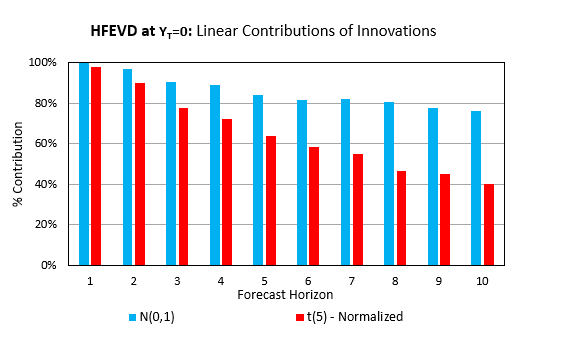}
	\caption{The cointributions of linear terms to the HFEVD for DAR(1) processes with N(0,1) innovations (blue bars) and t(5) innovations (red bars).}
	\label{fig:darfigure2}
\end{figure}

Next, an HFEVD analysis of both processes at horizon 2 with conditioning value $Y_T=0$ is shown in Figure 3, which provides a granular separation of contributions by the time horizon and by the degree of nonlinearity. The processes are approximated by a sum of four Hermite polynomial functions of structural innovations:
\begin{equation}\label{y2partial}
	y_{t+2} \approx c_{(1,0)}H_1(u_{t+1}) + c_{(0,1)}H_1(u_{t+2}) + c_{(1,1)}H_1(u_{t+1})H_1(u_{t+2}) + c_{(2,1)}H_2(u_{t+1})H_1(u_{t+2}),
\end{equation}
that capture the linear effects of $u_{t+1}$, the linear effects of $u_{t+2}$, the interaction of linear effects between $u_{t+1}$ and $u_{t+2}$, and the interaction of the quadratic effect of $u_{t+1}$ and the linear effect of $u_{t+2}$. For the process with N(0,1) innovations (left pie chart of Figure 3), this approximation sufficiently captures all the variation in $y_{t+2}$, as the percentage contribution of the four terms sum up to 100\%. The majority of the variation is explained by the marginal linear contributions of $u_{t+1}$ and $u_{t+2}$, accounting for 97\% (15\% + 82\%) of the conditional variance in $y_{t+2}$. However, 3\% of the remaining variation is due to the interaction term between the quadratic effect of $u_{t+1}$ and the linear effect of $u_{t+2}$. It is worth noting that there are no interacting linear effects between $u_{t+1}$ and $u_{t+2}$. For the model with t(5) innovations (right pie chart of Figure 3), the partial expansion only explains 97\% (77\%+12\%+8\%) of the variation in $y_{t+2}$ conditional on $y_T=0$. The remaining contributions are due to higher order effects, which is to be expected due to the non-Gaussianity of the structural innovations. The interaction term between the quadratic effect of $u_{t+1}$ and the linear effect of $u_{t+2}$ also plays a larger role in explaining the total variation. 

\begin{figure}[h!]
	\centering
	\includegraphics[width=1\linewidth]{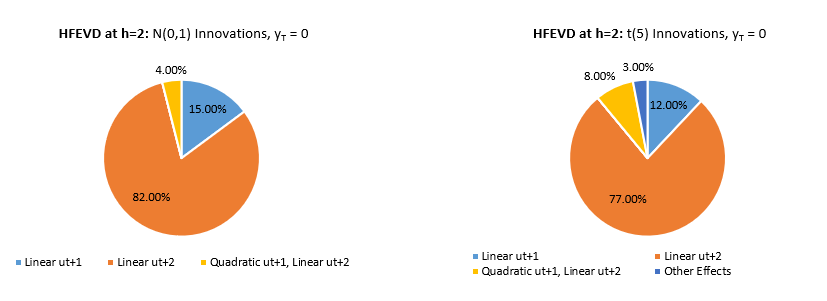}
	\caption{A focus on the HFEVD at horizon $h=2$ for the partial expansion \eqref{y2partial} of DAR(1) processes with N(0,1) innovations (left chart) and t(5) innovations (right chart).}
	\label{fig:darfigure3}
\end{figure}

\begin{figure}[h!]
	\centering
	\includegraphics[width=1\linewidth]{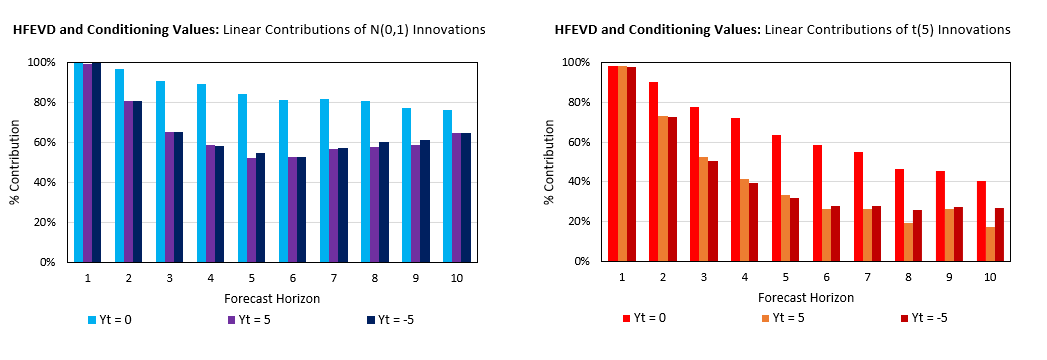}
	\caption{The HFEVD at different conditioning values $Y_T=0$, $Y_T=5$ and $Y_T=-5$ for the DAR(1) processes with N(0,1) innovations (left graph) and t(5) innovations (right graph)}
	\label{fig:darfigure4}
\end{figure}

The importance of the conditioning is shown in Figure 4, where the HFEVD is applied to both processes conditional on $Y_T=0$, $Y_T=5$ and $Y_T=-5$, respectively. Both graphs show that the decomposition can be sensitive to conditioning. In particular, at the mean value of the process $Y_T=0$, the linear contribution diminishes slowly as the horizon increases, while at more extreme values such as $Y_T=5$, or $Y_T=-5$, the decrease is much steeper. However, note that the HFEVD look similar for $Y_T=5$, or $Y_T=-5$. Hence, the decomposition analysis demonstrates that the DAR(1) exhibits symmetry with respect to the HFEVD under both Gaussian and t-distributed innovations.

\subsection{Stochastic Volatility Model}

The stochastic volatility models in \eqref{threshold} and \eqref{smootht} are supplied with the parameter values $a=0.5$, $a_1=0.7$, $a_2=0.4$, $\phi=0.6$ and $b=0.7$. This yields: 
\begin{equation}
	\begin{split}\label{thresholde}
		y_t& = 0.5 + 0.7y_{t-1} + \left[0.7\times\textbf{1}_{y_{t-1}\geq0} + 0.4 \times \textbf{1}_{y_{t-1}<0}\right]z_{t-1}+\exp(z_{t-1})\varepsilon_{1,t}, \\
		z_t& = 0.6 z_{t-1} + \varepsilon_{2,t},\\
	\end{split}
\end{equation} 
for the threshold transition and:
\begin{equation}\label{smoothe}
	\begin{split}
		y_t& = 0.7 +  0.7y_{t-1}+\left[0.7+\frac{-0.3}{1+\exp\left(-y_{t-1}\right)}\right]z_{t-1}   + \exp(z_{t-1})\varepsilon_{1,t}, \\
		z_t& = 0.5 z_{t-1} + \varepsilon_{2,t},\\
	\end{split}
\end{equation} 
for the smooth transition. Several exercises with the HFEVD are performed, to study the linear, specific quadratic and interaction contributions of innovations to the total variance conditional on $\underline{y_t}=(0,0)$:
\begin{enumerate}
	\item \textbf{Linear Contribution} for  $\varepsilon_{1,t+1},...,\varepsilon_{1,t+5}$:
 	\begin{equation*}
\sum_{i=1}^5	\mathbb{V}\left[C_{K_{1,i}}(\underline{y_t})H_1(\varepsilon_{1,t+i})\right],
\end{equation*}
where $K_{1,i}$ is the $10 \times 1$ vector associated with Hermite polynomial $H_1(\varepsilon_{2,t+i})$.

	\item \textbf{Linear Contribution} for $\varepsilon_{2,t+1},...,\varepsilon_{2,t+5}$:
 	\begin{equation*}
	\sum_{i=1}^5	\mathbb{V}\left[C_{K_{2,i}}(\underline{y_t})H_1(\varepsilon_{2,t+i})\right],
\end{equation*}
where $K_{2,i}$ is the $10 \times 1$ vector associated with Hermite polynomial $H_1(\varepsilon_{1,t+i})$.

	\item 	\textbf{Specific Quadratic Contribution} for $\varepsilon_{1,t+1},...,\varepsilon_{1,t+5}$:
\begin{equation*}
	\sum_{i=1}^5	\mathbb{V}\left[C_{K_{1,i}}(\underline{y_t})H_2(\varepsilon_{1,t+i})\right]
\end{equation*}
where $K_{1,i}$ is the $10 \times 1$ vector associated with Hermite polynomial $H_2(\varepsilon_{1,t+i})$.

	\item 	\textbf{Specific Quadratic Contribution} for $\varepsilon_{2,t+1},...,\varepsilon_{2,t+5}$:
\begin{equation*}
	\sum_{i=1}^5	\mathbb{V}\left[C_{K_{2,i}}(\underline{y_t})H_2(\varepsilon_{2,t+i})\right]
\end{equation*}
where $K_{2,i}$ is the $10 \times 1$ vector associated with Hermite polynomial $H_2(\varepsilon_{2,t+i})$.

	\item 	\textbf{Joint (Interaction) Linear Contribution:} $\varepsilon_{1,t+1},...,\varepsilon_{1,t+5}$ and  $\varepsilon_{2,t+1},...,\varepsilon_{2,t+5}$:
\begin{equation*}
	\sum_{i=1}^5	\mathbb{V}\left[C_{K_{i}}(\underline{y_t})H_1(\varepsilon_{1,t+i})H_1(\varepsilon_{2,t+i})\right]
\end{equation*}
where $K_{i}$ is the $10 \times 1$ vector associated with joint Hermite polynomial $H_1(\varepsilon_{1,t+i})H_1(\varepsilon_{2,t+i})$.

	\item 	\textbf{Joint (Interaction)} of \textbf{Linear Contribution} for  $\varepsilon_{1,t+1},...,\varepsilon_{1,t+5}$ and \textbf{Specific Quadratic Contribution} for $\varepsilon_{2,t+1},...,\varepsilon_{2,t+5}$:
\begin{equation*}
	\sum_{i=1}^5	\mathbb{V}\left[C_{K_{i}}(\underline{y_t})H_1(\varepsilon_{1,t+i})H_2(\varepsilon_{2,t+i})\right]
\end{equation*}
where $K_{i}$ is the $10 \times 1$ vector associated with joint Hermite polynomial $H_1(\varepsilon_{1,t+i})H_2(\varepsilon_{2,t+i})$.

	\item 	\textbf{Joint (Interaction)} of \textbf{Linear Contribution} for  $\varepsilon_{1,t+1},...,\varepsilon_{1,t+5}$ and \textbf{Specific Cubic Contribution} for $\varepsilon_{2,t+1},...,\varepsilon_{2,t+5}$:
\begin{equation*}
	\sum_{i=1}^5	\mathbb{V}\left[C_{K_{i}}(\underline{y_t})H_1(\varepsilon_{1,t+i})H_3(\varepsilon_{2,t+i})\right]
\end{equation*}
where $K_{i}$ is the $10 \times 1$ vector associated with joint Hermite polynomial $H_1(\varepsilon_{1,t+i})H_3(\varepsilon_{2,t+i})$.
\end{enumerate}

\begin{table}[h!]\centering
	\begin{tabular}{lcc}
		\hline
		\hline
		Effects & \multicolumn{2}{l}{Variance Contribution} \\
		  & \eqref{thresholde}& \eqref{smoothe} \\
		\hline
		\hline
		Linear Contributions of $\varepsilon_{1,t+1:t+5}$      &   21.53\%     &    20.82\%     \\
		\hline
		Linear Contributions of $\varepsilon_{2,t+1:t+5}$    &  3.38\%  & 2.15\%   \\      
		\hline
		Specific Quadratic Contributions of $\varepsilon_{1,t+1:t+5}$  &  0.0033\%   &  0.0034\%    \\    
		\hline
		Specific Quadratic Contributions of $\varepsilon_{2,t+1:t+5}$ &   0.0332\%   & 0.0227\%    \\        
		\hline
		Joint Interaction of Linear Contributions   & 21.96\%     &  21.39\%   \\    
	$\varepsilon_{1,t+1:t+5}$	and $\varepsilon_{2,t+1:t+5}$   &      &    \\    
		\hline
		Joint Interaction of Linear Contribution $\varepsilon_{1,t+1:t+5}$ & 10.82\%     &   10.88\%  \\    
		and Specific Quadratic Contribution of $\varepsilon_{2,t+1:t+5}$ &        &      \\    
		\hline
		Joint Interaction of Linear Contribution $\varepsilon_{1,t+1:t+5}$    &   3.60\%  &  3.78\%  \\    
		and Specific Cubic Contribution of $\varepsilon_{2,t+1:t+5}$ &  &         \\   
		\hline		
		\hline     
	\end{tabular}
	\caption{HFEVD Analysis for Models \eqref{thresholde} and \eqref{smoothe}.}
\end{table}

The results are summarized below in Table 1. As expected, the conditional variance contributions for both models are quite similar. This is not surprising, since they have a similar structure, with the only difference accounted for by the type of transition (threshold vs smooth). The degree of nonlinearity considered is at most up to order three for the threshold model. This accounts for $61.29\%$ ($\approx 21.53+3.38+21.96+10.82+3.60)$ of the total conditional variance, whereas for the smooth transition model, this accounts for $59.02\%$ ($\approx 20.82 + 2.15 + 21.39 + 10.88 + 3.78$) of the total conditional variance.
Second, the HFEVD can also be used to inform the practitioner on whether the use of a linear model (or linearization of a nonlinear model) is appropriate in the context of the data. For instance, the linear contribution each innovation component only accounts for $24.91\% (\approx 21.53+3.38)$ and $22.97\% (\approx 20.82 + 2.15 )$ of the variation in $Y_{t+5}$. Hence, the use of a linear model or approximation would misrepresent the trajectory for the underlying data generating process. Third, the HFEVD demonstrates that it is insufficient to include only marginal (specific) higher order terms; in this example, these specific quadratic contributions of the structural innovations account for almost zero in both models. This emphasizes the point of Isakin and Ngo (2020), who argue that interactions are indeed important for IRF and nonlinear FEVD analysis\footnote{As a secondary example, it is suggested in \cite{J2005}, equation 16, that a local projection (IRF) can be computed based on higher order polynomial terms, while explicitly omitting interactions ``as a matter of choice and parismony". This example shows that such a choice would be misguided.} Lastly, although these exercises have included 35 ($7 \times 5$) different Hermite polynomial terms, only 60\% of the total conditional variance has been accounted for. Indeed, since the Volterra expansion can potentially have an infinite amount of terms, a highly nonlinear process such as a model with threshold or smooth transitions can be comprised of many higher-order nonlinearities.

\section{Application to Credit Markets and Fiscal Policy}

The use of the HFEVD in an applied setting is demonstrated through the lens of the \cite{FRF2015} Threshold VAR (TVAR) framework (FRF2015 hereafter). The goal of their paper is to study how credit conditions can influence the transmission of fiscal policy shocks in the US macroeconomy. In particular, FRF2015 postulate that fiscal policy channels on aggregate output should be more effective under tighter credit conditions. Indeed, government spending boosts aggregate demand, driving increased household consumption by providing financial support to those who face borrowing constraints. Moreover, while expansionary fiscal policies often crowd-out private investment, they can mitigate financial limitations for firms, making it easier for them to grow when access to credit is limited. To investigate these claims empirically, FRF2015 propose a TVAR model on US quarterly data from 1984 to 2010. This section will first present the FRF2015 methodology, followed by an exposition of how the HFEVD can be used in their framework.

\subsection{The Model}

Suppose the dynamics of the macroeconomy are captured by a six-dimensional vector of time series $Y_t=(Y_{1,t},...,Y_{6,t})'$, containing measures of fiscal policy, output, public debt, inflation, monetary policy and credit conditions, respectively. The FFR2015 model is given by:
\begin{equation}\label{TVAR}
	\begin{split}
		Y_t &  = \begin{cases}
			A_1 Y_{t-1} + D_1\varepsilon_{t}, \ \ \ \textup{if} \ \ Y_{6,t-1} \geq r, \\
			A_2 Y_{t-1} + D_2\varepsilon_{t}, \ \ \ \textup{if} \ \ Y_{6,t-1} < r,\\ 
		\end{cases}\\
	\end{split}
\end{equation}
where the process $(Y_t)$ evolves according to a self-exciting TVAR(1) process [\cite{T1983}]. The two regimes depend on credit conditions in the previous period, $Y_{6,t-1}$, and the structural innovations $(\varepsilon_{t})$ are assumed to be i.i.d. standard Gaussian. Under ``tight" credit conditions characterized by $Y_{6,t-1} \geq r$ (where $r$ denotes the threshold value), the process follows a linear SVAR(1) model governed by coefficients $(A_2,D_2)$. Under ``ordinary" credit conditions such that $Y_{6,t-1} < r$, the economy follows a different linear SVAR(1) process, governed by coefficients $(A_1,D_1)$. \\

\begin{table}[h!]
	\begin{tabular}{p{3cm}|p{8.5cm}|p{3.5cm}}
			\hline 
				\hline 
		\textbf{Variable}       & \textbf{Data  }                                                                 & \textbf{Transformation}              \\
			\hline 
		\hline 
		Fiscal Policy     & Real Government Consumption and Gross Investment                       & Rate of Change                \\
			\hline 
		Output            & Real Gross Domestic Product (GDP)                                      & Rate of Change                \\
			\hline 
		Public Debt       & Ratio Between Government Debt and GDP                                  & First Difference              \\
			\hline 
		Inflation         & GDP Deflator                                                           & Logged Difference             \\
			\hline 
		Monetary Policy   & Effective Federal Funds Rate                                           & No Transformation             \\
			\hline 
		Credit Conditions & Spread Between BAA Corporate Bonds and 10-year Treasury Maturity Rates & MA(2) of the First Difference \\
			\hline 
				\hline 
	\end{tabular}
\caption{The variables, data and the transformations applied by FRF2015. }
\end{table}

Table 1 provides a summary of the variables and the transformations applied on the data (see Section 4 of their paper for a detailed discussion). The model is parametric and a two step estimation procedure is used:
\begin{enumerate}
	\item Conditional on a threshold value $r$, the reduced form model \eqref{TVAR} parameters $(A_1,A_2)$ and $(\Sigma_1,\Sigma_2)=(D_1D_1',D_2D_2')$ can be estimated using OLS. To select the optimal choice of $r$, a grid of values is considered and the one that maximizes the Akaike Information Criterion (AIC) is chosen. 
	\item The structural parameters $(D_1,D_2)$ cannot be identified from the first step, and this problem is analogous to the linear SVAR(1) model. FRF2015 propose a recursive ordering identification scheme where a fiscal policy shocks influences the variables in the following order: [1] Fiscal Policy [2] Output [3] Public Debt [4] Inflation [5] Monetary Policy [6] Credit Conditions. Then, a unique lower triangular Cholesky decomposition for the covariance matrix of each regime is performed to derive the estimates of $D_1$ and $D_2$ \footnote{Recall that ``this mechanical solution does not make economic sense without a plausible economic interpretation for the recursive ordering" [see \cite{KL2017}, Section 8.2 for a deep discussion of the limitations in recursive systems for impulse analysis].}.
\end{enumerate}
The effects of a fiscal policy shock (or its contribution) can then be studied by computing nonlinear IRFs (or nonlinear FEVDs) on its structural innovation, that is, $\varepsilon_{1,t}$. 

\subsection{HFEVD Analysis}

The HFEVD can be used to assess the nonlinearity of the model and to study the contributions of fiscal policy shocks. This subsection will focus on on six distinct periods in history, representing the starts and ends of three recessions in the United States as defined by the NBER. Table 3 summarizes these dates and indicates whether the observed history is in an ``ordinary" credit regime ($r < 0.17$) or ``tight" credit regime ($r \geq 0.17$). 
\begin{table}[h!]
	\centering
	\begin{tabular}{lll}
		\hline
		\hline
		Recession            & Start            & End          \\
		\hline
		\hline
		Early 90's Recession & 1990Q3           & 1991Q1       \\
		& \textcolor{blue}{Ordinary (-0.12)} & \textcolor{red}{Tight (0.31)} \\
		\hline
		Early 00's Recession & 2001Q2           & 2001Q4       \\
		& \textcolor{red}{Tight (0.20)}     & \textcolor{red}{Tight (0.18)} \\
		\hline
		Great Recession      & 2007Q3           & 2009Q2       \\
		& \textcolor{blue}{Ordinary (-0.03)} & \textcolor{red}{Tight (1.52)}\\
		\hline
		\hline
	\end{tabular}
	\caption{The histories considered for the HFEVD analysis.}
\end{table}

\subsubsection{Linear Contributions of All Innovations}

A first question of practical significance is how much variation in $Y_{t+h}$ is accounted for by linear contribution of all the innovations, conditional on vector $\underline{y_t}$. This is equivalent of performing HFEVD analysis on a linear combination of all first-order Hermite polynomials for the structural innovations (i.e. the partial Wold representation of the process at horizon $h$). Indeed, if the model is truly nonlinear, one would expect to see varying results for different values of $\underline{y_t}$ (i.e. dependent on history) and contributions of less than 100\% (since 100\% would imply linearity). In Figure 5, each plot contains six lines, which represent the linear contribution of all structural innovations on the variance of a particular variable, conditional on one of the historical periods. For example, the light blue line in the top left graph represents the linear contribution of structural innovations on the variance of credit conditions, conditional on the data observed in 1990 Q3. Note that in all the graphs, as the horizon increases, the contributions ``converge" to a particular value. Indeed, since the data was treated by FRF2015 to ensure its stationarity and the variance will approach its long run value as the forecast horizon tends to infinity. \\

\begin{figure}[h!]
	\centering
	\includegraphics[width=1\linewidth]{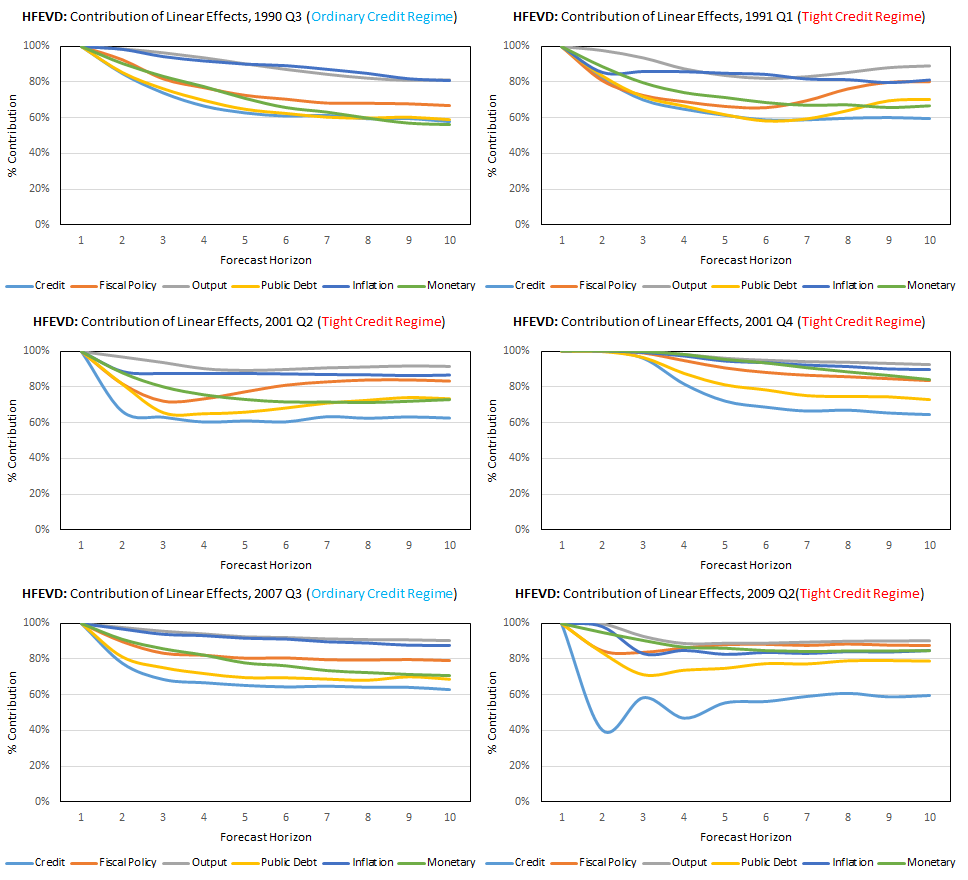}
	\caption{The linear contribution to the variance of each variable in the TVAR, conditional on histories: 1991 Q3 [Top left], 1991 Q1 [Top right], 2001 Q2 [Middle left], 2001 Q4 [Middle right], 2007 Q3 [Bottom left] and 2009 Q2 [Bottom right].}
	\label{fig:application-1}
\end{figure}

 For the top left graph, the HFEVD is performed conditional on the observation at 1990Q3, which was in an ordinary credit regime. Both output and inflation exhibit fairly linear behaviour, with the linear contribution approaching 80\% of the total variation in the long run. However, fiscal policy, credit, monetary policy and public debt exhibit more nonlinear behaviour, with the contribution of the linear structural innovations approaching 60\% in the long-run. At the end of the Early 90's recession in 1991Q1, there is a change in credit conditions to a tighter regime. The linear behaviour of output, fiscal policy, public debt, and inflation seemingly increase after forecast horizon 5, and converge to a new long-run value. However, the linear contribution to credit and inflation remain the same.  In the second row, the results for the Early 00's recession is shown. Output, inflation and fiscal policy tend to exhibit linear behaviour, while public debt, monetary policy and credit behave more nonlinearly. Although the long run linear contribution of the innovations converges to similar values in both 2001 Q2 and 2001 Q4, there are different short-run effects. In particular, for the latter time period, the linear contribution of all structural innovations is 100\% for the first three horizons. This suggests that the economy stayed linear in one regime for three forecast horizons before nonlinear dynamics returned. The results for the '08 financial crisis are shown in the last two graphs of Figure 5. The start of the crisis was characterized by an ordinary credit regime\footnote{This is an ordinary credit regime in the framework of this model. However, at this time, there was a decrease of the quality of newly issued credits. This variable has not been introduced in the model and it is important to keep in mind that the contributions may also depend on the list of variables included in the system.}, with the linear behavior of all six variables similar to some of the other periods above. However, at the conclusion of the crisis, it was evident that there were significant short-run and long-run changes on how the structural innovations contributed linearly to the variation in the economy. For credit in particular, there is a significant decline in linear contributions at horizon 2 for 2009Q2, that persists until horizon 6 before returning to the long-run value. On the other hand, inflation and public debt seem to converge towards different values between 2007Q3 and 2009Q2, with a higher proportion of linear contributions for the latter history.
 
 \subsubsection{Contributions of Fiscal Policy Shocks on Output}
 
To address original research question of FRF2015, the HFEVD can be used in this context to understand how  fiscal policy shocks influence economic output. However, it is important to note that FRF2015 uses the Generalized Impulse Response Function (GIRF) to characterize and compute nonlinear IRFs. As mentioned in Section 4.5 and Appendix B, there are issues with such a definition and the interpretation of the GIRF is not compatible with the HFEVD. Hence, it is important to keep in mind that the results of the HFEVD here are associated with shocks according to the perturbed and baseline paths in \eqref{nonlinear_irf_paths}, and the EIRF in \eqref{EIRF}. 

\begin{figure}[h!]
	\centering
	\includegraphics[width=1\linewidth]{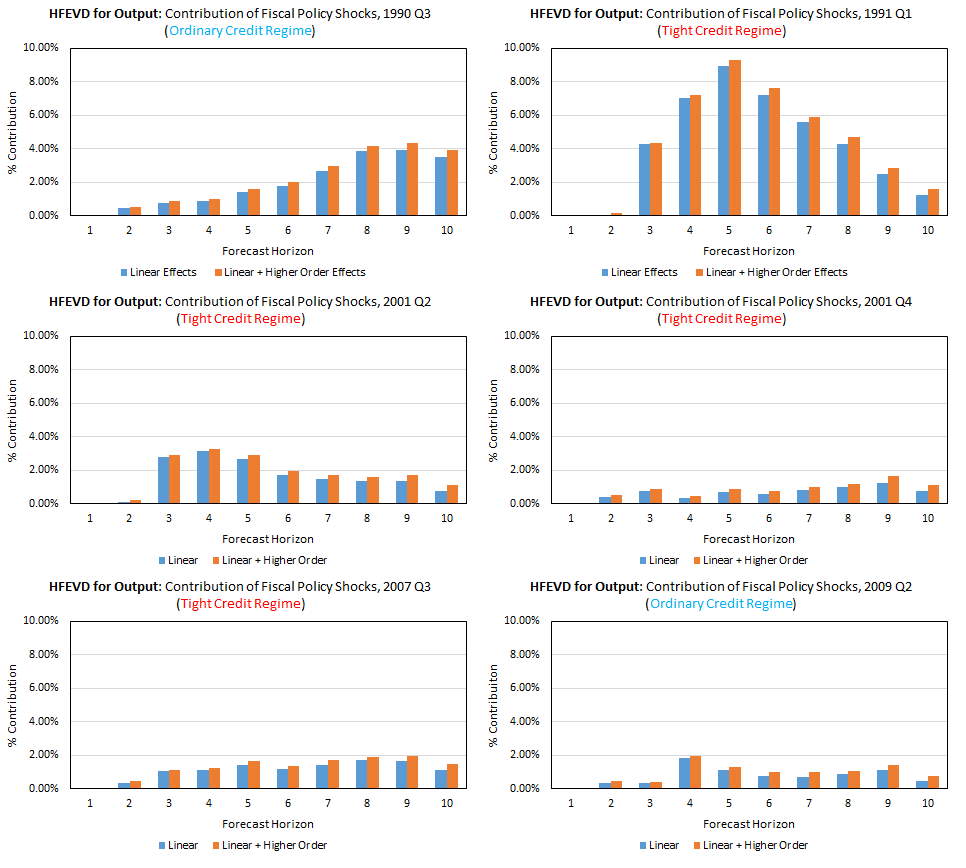}
	\caption{The contribution of marginal effects of fiscal policy shocks to the variance of output, conditional on histories: 1991 Q3 [Top left], 1991 Q1 [Top right], 2001 Q2 [Middle left], 2001 Q4 [Middle right], 2007 Q3 [Bottom left] and 2009 Q2 [Bottom right].}
	\label{fig:application-2}
\end{figure}

Each graph in Figure 6 contains the marginal contribution of fiscal policy shocks on variance of economic output conditional on the histories in Table 3. There are three insights from these results. First, there is little reason to suggest that the marginal contribution of fiscal policy shocks are nonlinear. Indeed, each graph shows that the linear contribution and the linear + specific higher order contribution (up to degree 5) are not very different\footnote{However, this says nothing about interactions between fiscal policy shocks with shocks on the other innovations. Under recursive ordering, only the fiscal policy shock has economic meaning, so the analysis of interactions are suppressed in this exposition.} Second, \cite{FRF2015} ``find that the responses of output to fiscal policies significantly change according to the state of credit markets" (Section 1). Indeed, from the start to the end of the Early 90's recession, there is a shift from an ordinary credit regime to a tightening of credit markets. The first two graphs of Figure 6 show that the contribution of fiscal policy shocks on the variance of output is higher in most forecast horizons conditional on 1991 Q1, in agreement with the statement made by FRF2015. Similarly, for the Great Recession, the contribution of fiscal policy shocks seem to have higher contribution on output in 2007 Q3 under a tighter credit regime compared to the ordinary regime in 2009 Q2. On the other hand, the analysis of the Early 2000's crisis dampens their claim. In particular, both the start and end of this recession are characterized by tight credit regimes with similar BAA spreads (0.20 and 0.18 for 2001 Q2 and 2001 Q4, respectively). However, the influence of fiscal policy shocks on the variance of output is lower conditional on the observed history at 2001 Q4. This contradiction with FRF2015 can be explained by the type of conditioning done in their paper. For the GIRF analysis, FRF2015 conditions only on the histories of the credit variable, but average out the histories of all other variables. This means they do not condition on specific dates, but rather average out the impact of fiscal policy shocks on several sets of dates that fit under ``tight" credit regimes (or ``ordinary" credit regimes). For instance, the results for 1991Q1, 2001Q2, 2001Q4, and 2007Q3 are all tight credit regimes, so their histories would be averaged out to generate a single result. However, even if two histories are within the same regime, there can be very different HFEVDs. As such, there can be misleading insights resulting from averaging out histories.

\section{Conclusion}

This paper explores a new method for FEVD analysis in nonlinear SVAR models called the HFEVD. It exploits the properties of Hermite polynomials and represents the trajectory of a Markov processes written on Gaussian innovations using a Hermite polynomial expansion. This representation facilitates the construction of an HFEVD since each term is orthogonal under the standard Gaussian density, leading to a deocmposition of effects by time horizon, by components of the innovations and by the degree of nonlinearity. A link between the terms in the HFEVD and the impact multipliers of the nonlinear IRF is also established. This extends an important property of FEVDs that also exists in the linear framework. Simulation exercises are provided to demonstrate how the HFEVD can be implemented in practice for both parametric models written on Gaussian innovations, or semiparametric models with cross-sectionally independent innovations that can be normalized to Gaussian. An application to fiscal policy shocks and credit markets under the lens of the FRF2015 TVAR framework is also provided. \\

The HFEVD has several useful applications. First, it can be used by policymakers to gauge the importance of innovations in a wide range of nonlinear dynamic models, including the ones that were not discussed in this paper. The granular view of effects allow practioners to assess the properties of shocks in their models that are obscured by existing methods in the literature. Secondly, the HFEVD is an effective tool for model appraisal. For instance, it can be used to evaluate the consequences of linearization for nonlinear models or in determining the feasibility of more complex specifications.

\newpage

\bibliography{references.bib}

\newpage

\appendix 

\section{Proofs}

This appendix provides the proofs of the key results in the main text.

\subsection{Proof of Lemma 3}

A Taylor expansion of the perturbed path $Y_{t+1}(\delta)$ around $\delta=0$ yields:
\begin{equation}
Y_{t+1}(\delta)=	g(\underline{y_t},\varepsilon_{t+1}+\delta) = g(\underline{y_t},\varepsilon_{t+1}) + \frac{\delta}{1!} \frac{d}{d\varepsilon_{t+1}}g(\underline{y_t},\varepsilon_{t+1})+\frac{\delta^2}{2!}\frac{d^2}{d\varepsilon_{t+1}^2}g(\underline{y_t},\varepsilon_{t+1})+...,
\end{equation}
which implies:
\begin{equation}
		g(\underline{y_t},\varepsilon_{t+1}+\delta) - g(\underline{y_t},\varepsilon_{t+1}) = \sum_{k=1}^\infty 
	\frac{\delta^k}{k!} \frac{d^k}{d\varepsilon_{t+1}^k}g(\underline{y_t},\varepsilon_{t+1}).
\end{equation}
The desired result is obtained by taking expectations conditional on $\underline{y_t}$:
\begin{equation}
	\mathbb{E}\left[g(\underline{y_t},\varepsilon_{t+1}+\delta) - g(\underline{y_t},\varepsilon_{t+1})|\underline{y_t}\right]=\mathbb{E}[Y_{t+1}(\delta) - 	Y_{t+1}|\underline{y_t}] = \sum_{k=1}^\infty \frac{\delta^k}{k!}\mathbb{E}\left[\frac{d^k}{d\varepsilon_{t+1}^k}g(\underline{y_t},\varepsilon_{t+1})\right].
\end{equation}
\qed

\subsection{Proof of Proposition 4}

Starting from Proposition 2, each term in the HFEVD for $k \geq 1$ can be written as: 
\begin{equation}
	\begin{split}
		c_k(\underline{y}_t)(k!)& =\mathbb{E}[g(\underline{y}_t,\varepsilon_{t+1})H_k(\varepsilon_{t+1})]= \int_{-\infty}^\infty g(\underline{y}_t,\varepsilon)H_k(\varepsilon)\phi(\varepsilon) d\varepsilon. \\
	\end{split}
\end{equation}
From the definition of Hermite polynomial in \eqref{hermite_uni}: 
\begin{equation}
\begin{split}\label{RFhu}
		&H_k(\varepsilon)=(-1)^k\exp\left(\frac{\varepsilon^2}{2}\right)\frac{d^k}{d\varepsilon^k}\left[\exp\left(-\frac{\varepsilon^2}{2}\right)\right] \\
		\iff  &  H_k(\varepsilon)\exp\left(-\frac{\varepsilon^2}{2}\right)=(-1)^k\frac{d^k}{d\varepsilon^k}\left[\exp\left(-\frac{\varepsilon^2}{2}\right)\right]  \\
		\iff   & H_k(\varepsilon)\phi(\varepsilon)=(-1)^k\frac{d^k}{d\varepsilon^k}\phi(\varepsilon) \\
\end{split}
\end{equation}
This implies:
\begin{equation}\label{a5}
	\begin{split}
		\int_{-\infty}^\infty g(\underline{y}_t,\varepsilon)H_k(\varepsilon)\phi(\varepsilon) d\varepsilon =& (-1)^k\int_{-\infty}^\infty g(\underline{y}_t,\varepsilon)\frac{d^k}{d\varepsilon^k}\phi(\varepsilon) d\varepsilon \\
		= &  (-1)^k \left[g(\underline{y}_t,\varepsilon)\frac{d^k}{d\varepsilon^k}\phi(\varepsilon) \right]_{-\infty}^{+\infty}-(-1)^k \int_{-\infty}^\infty  \frac{d}{d\varepsilon} g(\underline{y}_t,\varepsilon)\frac{d^{k-1}}{d\varepsilon^{k-1}}\phi(\varepsilon) d\varepsilon \\
		& \text{(Using integration by parts)}\\
		= & 0 + (-1)^{k+1} \int_{-\infty}^\infty  \frac{d^k}{d\varepsilon^k} g(\underline{y}_t,\varepsilon)\phi(\varepsilon) d\varepsilon \\
		= & (-1)^{k+1}\mathbb{E}\left[\frac{d^k}{d\varepsilon^k} g(\underline{y}_t,\varepsilon)\right].\\
	\end{split}
\end{equation}
Squaring results in:
\begin{equation}
	\left[c_k(\underline{y}_t)(k!)\right]^2=\mathbb{E}\left[\frac{d^k}{d\varepsilon^k} g(\underline{y}_t,\varepsilon)\right]^2,
\end{equation}
as required. \qed

\subsection{Proof of Proposition 5}

First, Lemma 3 is extended to this bivariate setting. Without loss of generality, suppose $\varepsilon=(\varepsilon_1,\varepsilon_2)$ is a bivariate Gaussian innovation with shock magnitude $\delta=(\delta_1,\delta_2)$. The baseline and perturbed paths can be written in simplified notation as:
\begin{equation}
	\begin{split}
		y & = g(\varepsilon),\\
		y(\delta) & = g(\varepsilon+\delta).
	\end{split}
\end{equation}
A Taylor expansion of $\mathbb{E}\left[g(\varepsilon+\delta)\right]$ around $0$ yields: 
\begin{equation}
	\begin{split}
		\mathbb{E}\left[g(\varepsilon+\delta)\right] & =\mathbb{E}\left[g(\varepsilon)\right] + \frac{\delta_1}{1!}\mathbb{E}\left[\frac{\partial g(\varepsilon)}{\partial \varepsilon_1}\right]+  \frac{\delta_2}{1!}\mathbb{E}\left[\frac{\partial g(\varepsilon)}{\partial \varepsilon_2}\right]+\frac{\delta_1\delta_2}{1!1!}\mathbb{E}\left[\frac{\partial g(\varepsilon)}{\partial \varepsilon_1\partial\varepsilon_2}\right]+ ...\\
		& =\mathbb{E}\left[g(\varepsilon)\right]+ \sum_{\substack{K \in \mathbb{N}^2 \\ \textbf{K}\neq 0}}\frac{\delta_1^{k_1}\delta_2^{k_2}}{k_1!k_2!}\mathbb{E}\left[\frac{\partial^{(k_1+k_2)}}{\partial\varepsilon_1^{k_1}\partial\varepsilon_2^{k_2}}g(\varepsilon)\right],
	\end{split}
\end{equation}
where $K=(k_1,k_2) \in \mathbb{N}^2$ captures the degree of differentiation in the Taylor expansion. Under the simplified notation, $EIRF(\delta) = \mathbb{E}\left[g(\varepsilon+\delta)\right]-\mathbb{E}\left[g(\varepsilon)\right]$, and we get the desired result. \\

\noindent Starting from Proposition 3, each term in the HFEVD for $k \geq 1$ can be written as: 
\begin{equation}
	\begin{split}
		C_K(\underline{y_t})(k_1!k_2!) &= \mathbb{E}\left[ g(y_t,\varepsilon_{1,t+1},\varepsilon_{2,t+1})H_{K}(\varepsilon_{1,t+1},\varepsilon_{2,t+1})\right]\\
		& = \mathbb{E}\left[ g(y_t,\varepsilon_{1,t+1},\varepsilon_{2,t+1})H_{k_1}(\varepsilon_{1,t+1})H_{k_2}(\varepsilon_{2,t+1})\right]\\
		& = \int_{-\infty}^\infty\int_{-\infty}^\infty g(\underline{y}_t,\varepsilon_1,\varepsilon_2)H_{k_1}(\varepsilon_{1})H_{k_2}(\varepsilon_{2})\phi_1(\varepsilon_1)\phi_2(\varepsilon_2)d\varepsilon_1d\varepsilon_2. \\
	\end{split}
\end{equation}
As in \eqref{a5}, for $\varepsilon_1$ and $\varepsilon_2$ we have:  
\begin{equation}\label{RFhm}
	\begin{split}
H_{k_1}(\varepsilon)\phi(\varepsilon_1)& =(-1)^{k_1}\frac{d^{k_1}}{d\varepsilon_1^{k_1}}\phi(\varepsilon_1), \\
H_{k_2}(\varepsilon)\phi(\varepsilon_2)& =(-1)^{k_2}\frac{d^{k_2}}{d\varepsilon_2^{k_2}}\phi(\varepsilon_2).\\
	\end{split}
\end{equation}
Hence:
\begin{equation}
	\begin{split}
		C_K(\underline{y_t})(k_1!k_2!) &= \int_{-\infty}^\infty\int_{-\infty}^\infty g(\underline{y}_t,\varepsilon_1,\varepsilon_2)(-1)^{k_1}\frac{d^{k_1}}{d\varepsilon_1^{k_1}}\phi(\varepsilon_1)(-1)^{k_2}\frac{d^{k_2}}{d\varepsilon_2^{k_2}}\phi(\varepsilon_2)d\varepsilon_1d\varepsilon_2. \\
	\end{split}
\end{equation}
Then, integration by parts can be done twice - first with respect to $\varepsilon_1$, then with respect to $\varepsilon_2$. This implies:
\begin{equation}
	\begin{split}
& \int_{-\infty}^\infty\int_{-\infty}^\infty g(\underline{y}_t,\varepsilon_1,\varepsilon_2)(-1)^{k_1}\frac{d^{k_1}}{d\varepsilon_1^{k_1}}\phi(\varepsilon_1)(-1)^{k_2}\frac{d^{k_2}}{d\varepsilon_2^{k_2}}\phi(\varepsilon_2)d\varepsilon_1d\varepsilon_2 \\
= & \int_{-\infty}^\infty \left[\int_{-\infty}^\infty g(\underline{y}_t,\varepsilon_1,\varepsilon_2)(-1)^{k_1}\frac{d^{k_1}}{d\varepsilon_1^{k_1}}\phi(\varepsilon_1)d\varepsilon_1\right](-1)^{k_2}\frac{d^{k_2}}{d\varepsilon_2^{k_2}}\phi(\varepsilon_2)d\varepsilon_2\\
= & \int_{-\infty}^\infty \left[(-1)^{k_1+1} \int_{-\infty}^\infty  \frac{d^{k_1}}{d\varepsilon_1^{k_1}} g(\underline{y}_t,\varepsilon_1,\varepsilon_2)\phi(\varepsilon_1) d\varepsilon_1\right](-1)^{k_2}\frac{d^{k_2}}{d\varepsilon_2^{k_2}}\phi(\varepsilon_2)d\varepsilon_2\\
= & (-1)^{k_1+1}\int_{-\infty}^\infty  \left[\int_{-\infty}^\infty  \frac{d^{k_1}}{d\varepsilon_1^{k_1}} g(\underline{y}_t,\varepsilon_1,\varepsilon_2)(-1)^{k_2}\frac{d^{k_2}}{d\varepsilon_2^{k_2}}\phi(\varepsilon_2)d\varepsilon_2\right]\phi(\varepsilon_1) d\varepsilon_1\\
= & (-1)^{k_1+1}(-1)^{k_2+1}\int_{-\infty}^\infty \int_{-\infty}^\infty  \frac{d^{k_1}}{d\varepsilon_1^{k_1}}\frac{d^{k_2}}{d\varepsilon_2^{k_2}} g(\underline{y}_t,\varepsilon_1,\varepsilon_2)\phi(\varepsilon_1)\phi(\varepsilon_2) d\varepsilon_1d\varepsilon_2\\
= &  (-1)^{k_1+1}(-1)^{k_2+1} \mathbb{E}\left[\frac{\partial^{(k_1+k_2)}}{\partial\varepsilon_{1,t+1}^{k_1}\partial \varepsilon_{2,t+1}^{k_2}}g(\underline{y_t},\varepsilon_{1,t+1},\varepsilon_{2,t+1})\right]
	\end{split}
\end{equation}
Hence by taking the square:
\begin{equation}
	\begin{split}
		  & \left[C_K(\underline{y_t})C_K(\underline{y_t})'(k_1!k_2!)^2\right]\\
		= &\mathbb{E}\left[\frac{\partial^{(k_1+k_2)}}{\partial\varepsilon_{1,t+1}^{k_1}\partial \varepsilon_{2,t+1}^{k_2}}g(\underline{y_t},\varepsilon_{1,t+1},\varepsilon_{2,t+1})\right]\mathbb{E}\left[\frac{\partial^{(k_1+k_2)}}{\partial\varepsilon_{1,t+1}^{k_1}\partial \varepsilon_{2,t+1}^{k_2}}g(\underline{y_t},\varepsilon_{1,t+1},\varepsilon_{2,t+1})\right]'
	\end{split}
\end{equation}
which is the intended result. \qed

\section{Various Definitions of the Nonlinear EIRF}

This appendix compares the various definitions of ``shocks" proposed in the nonlinear IRF literature by considering their Taylor expansions around an infinitesimal shock. For exposition, consider the one-dimensional case and a one-time transitory shock at horizon 1. 

\begin{enumerate}
	\item \textbf{Nonlinear Structural IRF \eqref{EIRF}:} A shock is characterized by $\varepsilon_{t+1}+\delta$ on impact, and all other structural innovations follow their natural trajectory [This definition is used in \cite{GJ2005}, Section 3.2, \cite{GEA2024}, Definition 1, \cite{RS2021} with a modified definition of the GIRF in Section 7.3.]. The EIRF is given by:
	\begin{equation*}
		EIRF(\delta,\underline{y_t}) =\mathbb{E}\left[g(\underline{y_t},\varepsilon_{t+1}+\delta)-g(\underline{y_t},\varepsilon_{t+1})|\underline{y_t}\right].
	\end{equation*}
By Lemma 3 we get: 
	\begin{equation}\label{t_eirf}
			\mathbb{E}\left[g(\underline{y_t},\varepsilon_{t+1}+\delta)-g(\underline{y_t},\varepsilon_{t+1})\right]= \delta \mathbb{E}\left[\frac{dg(\underline{y_t},\varepsilon_{t+1})}{d\varepsilon_{t+1}}\right] + \frac{\delta^2}{2!}\mathbb{E}\left[\frac{d^2g(\underline{y_t},\varepsilon_{t+1})}{d\varepsilon_{t+1}^2}\right] + ...
	\end{equation}\\

	\item \textbf{Koop, Pesaran and Potter (1996) GIRF:}
	 A shock is characterized by $\varepsilon=\delta$ on impact, and all other structural innovations follow their natural trajectory [See also \cite{P2000}, example 4.2.1., \cite{LN2016}, \cite{P2015}, Chapter 24.5, \cite{KL2017}, Chapter 18.2.2]. The GIRF is given by:
	\begin{equation*}
		GIRF(\delta,\underline{y_t}) =\mathbb{E}\left[g(\underline{y_t},\delta)-g(\underline{y_t},\varepsilon_{t+1})|\underline{y_t}\right],
	\end{equation*}
	and a Taylor expansion of $g(\varepsilon_{t+1},\underline{y_t})$ around $\delta=0$ yields:
	\begin{equation*}
		g(\underline{y_t},\delta)= g(\underline{y_t},0) + \delta \frac{dg(\underline{y_t},\varepsilon_{t+1})}{d\varepsilon_{t+1}}\bigg \vert_{\varepsilon_{t+1}=0}+ \frac{\delta^2}{2!}\frac{d^2g(\underline{y_t},\varepsilon_{t+1})}{d\varepsilon_{t+1}^2}\bigg \vert_{\varepsilon_{t+1}=0} + ... ,
	\end{equation*}
	Hence we get: 
		\begin{equation}\label{t_girf}
		GIRF(\delta,\underline{y_t})= g(\underline{y_t},0)-\mathbb{E}\left[g(\underline{y_t},0)\right] + \delta \mathbb{E}\left[\frac{dg(\underline{y_t},0)}{d\varepsilon_{t+1}}\right]+ \frac{\delta^2}{2!}\mathbb{E}\left[\frac{d^2g(\underline{y_t},0)}{d\varepsilon_{t+1}^2}\right] + ... 
	\end{equation}
	\item \textbf{MIT Shock:} 
	A shock is characterized by $\delta$ on impact, and all other structural innovations are set directly equal to 0.  The EIRF is given by:
		\begin{equation*}
		EIRF(\delta) =\mathbb{E}\left[g(\underline{y_t},\delta)-g(\underline{y_t},0)|\underline{y_t}\right],
	\end{equation*}
	and a Taylor expansion of $g(\underline{y_t},\delta)$ around $\delta=0$ yields:
\begin{equation*}
	g(\underline{y_t},\delta)= g(\underline{y_t},0) + \delta \frac{dg(\underline{y_t},0)}{d\varepsilon_{t+1}}+ \frac{\delta^2}{2!}\frac{d^2g(0)}{d\varepsilon_{t+1}^2} + ... ,
\end{equation*}
Hence we get: 
\begin{equation}\label{t_mit}
		\mathbb{E}\left[g(\underline{y_t},\delta)-g(\underline{y_t},0)\right]=  \delta \frac{dg(\underline{y_t},0)}{d\varepsilon_{t+1}}+ \frac{\delta^2}{2!}\frac{d^2g(\varepsilon)}{d\varepsilon_{t+1}^2} + ... 
\end{equation}
\end{enumerate}

As discussed in Lemma 3, each term in the Taylor expansion is the product of the (scaled) magnitude and the expected impact multiplier. In fact, each of the expansions in \eqref{t_eirf}, \eqref{t_girf}  and \eqref{t_mit} have a similar form, but there are two key differences in the latter interpretations. For the \cite{KPP1996} definition, an infinitesimal shock will have in general, a significant impact, since $g(\underline{y_t},0)-\mathbb{E}\left[g(\underline{y_t},\varepsilon_{t+1})\right]\neq0$. This effect disappears only if the function $g$ is linear in $\varepsilon$, or at horizon 1 for nonlinear models (in $\underline{y_t}$) that are linear in $\varepsilon$ (but no longer after horizon 1). For the MIT shock, the impact multipliers are evaluated at 0 which neglects the uncertainty in $\varepsilon$ for the higher degree coefficients. In particular, $\mathbb{E}\left[\frac{d^2g(\varepsilon)}{d\varepsilon^2}\right]\neq \frac{d^2g(\mathbb{E}(\varepsilon))}{d\varepsilon^2} = \frac{d^2g(0)}{d\varepsilon^2}$. \\

\section{Estimation Risk and the HFEVD}

An area of FEVD analysis that has seen limited discussion is the uncertainty of these measures due to estimation error. In practice, the function $g$ is not known and needs to be estimated. In a parametric model for example, let: 
\begin{equation*}
	y_t=g(\underline{y_{t-1}},\varepsilon_t;\theta_0),
\end{equation*}
where $\theta$ is the parameter set. Suppose that the model is well specified with $\theta_0$ as the true parameter value and that there is a consistent an asymptotically normal estimator $\hat{\theta}_T$ of $\theta_0$. Then: 
\begin{equation}\label{a_normality}
	\sqrt{T}(\hat{\theta}_T - \theta_0 ) \sim N(0,\Sigma),
\end{equation}
as $T$ tends to infinity. For exposition, consider the univariate case, and look at the first-order coefficients of the Hermite polynomial expansion. 
\begin{equation}
c_k(\underline{y}_t;\theta_0)(k!)=\mathbb{E}\left[\frac{d^k}{d\varepsilon^k} g(\underline{y}_t,\varepsilon;\theta_0)\right],
\end{equation}
The corresponding estimated coefficient would be:
\begin{equation*}
	c_k(\underline{y}_t;\hat{\theta}_T)(k!)= \mathbb{E}\left[\frac{d^k}{d\varepsilon^k} g(\underline{y}_t,\varepsilon;\hat{\theta}_T)\right].
\end{equation*}
Using the integration by parts formula we obtain: 
\begin{equation*}
	\begin{split}
		\sqrt{T}\left[c_k(\underline{y}_t;\hat{\theta}_T)(k!)-c_k(\underline{y}_t;\theta_0)(k!)\right]=\sqrt{T}\left[\frac{d^k}{d\varepsilon^k} g(\underline{y}_t,\varepsilon;\hat{\theta}_T)-\frac{d^k}{d\varepsilon^k} g(\underline{y}_t,\varepsilon;\theta_0)\right] . \\
	\end{split}
\end{equation*}
Applying the multivariate delta method on \eqref{a_normality}, we know that the above expression converges in distribution to:
\begin{equation}
	N\left(0,\mathbb{E}\left[\frac{\partial^{k+1}}{\partial\varepsilon^k\partial\theta}g(y_t,\varepsilon,\theta_0)\right]\ \Sigma \ \mathbb{E}\left[\frac{\partial^{k+1}}{\partial\varepsilon^k\partial\theta}g(y_t,\varepsilon,\theta_0)\right]'\right).
\end{equation}

\section{Orthogonal FEVD with Non-Gaussian Innovations}

In some contexts, a practitioner may be interested in performing FEVD analysis on a parametric model with non-Gaussian innovations, rather than applying the HFEVD on innovations that are standardized to Gaussian. The aim of this appendix is to discuss this possibility and to extend the strategy proposed in the main text into a generalized method called the Orthogonal FEVD (OFEVD). The major caveat however, is that the link between the EIRF and terms of the OFEVD no longer holds. \\

Recall the set-up presented in \eqref{NLAR_alt} given by:
\begin{equation}
	Y_t = g^*(\underline{Y_{t-1}},u_t),
\end{equation}
where $g^*$ is invertible in $u_t=(u_{1,t},...,u_{n,t})$, $u_t$ is a strong white noise with independent components a p.d.f. $f_u$ (with corresponding c.d.f $F_u$), which is not necessarily Gaussian. It is important to reiterate that the identification issue that arises when the dimension of $Y_t$ is greater than 1. The future trajectory of this process can be generated by repeated substitution, similar to Section 3.2. At horizon 1 the trajectory is given by:
\begin{equation}
	Y_{t+1} = g^*(\underline{Y_{t}},u_{t+1}), 
\end{equation}
and through repeated substitution, the process at horizon $h$ becomes:
\begin{equation}\label{ftng}
	Y_{t+h} = g^{*(h)}(\underline{Y_{t}},u_{t+1:t+h}), 
\end{equation}
where $u_{t+1:t+h}=(u_{1,t+1},...,u_{1,t+h},...,u_{n,t+1},...,u_{n,t+h})'$ and  $g^{*(h)}(\cdot)$ is the function $g^*$ applied recursively $h$ times. Thus, the future trajectory is a function of history $\underline{Y}_t$ and $n \times h$ independent random variables.

\subsection{Generating Orthogonal Polynomials}

To proceed with OFEVD construction, the first task is to generate a sequence of orthogonal polynomials with respect to the densities of the innovations to facilitate the Volterra type representation of the future trajectory. Recall from Section 4.1.1, a sequence of polynomials $P_k(U)$ of degree $k$ are orthogonal on the $A$, the support of $U$, with respect to the weight function $w(U)$ if they satisfy:

\begin{equation}\label{ortho_poly}
	\int_AP_k(u)P_m(u)w(u)du = 0,
\end{equation} 
where $k \neq m$ and $k,m \in \mathbb{N}$. Orthogonal polynomials appear naturally for some special distributions, such as for Gaussian, Gamma, Uniform or Beta innovations. In these standard cases, the sequence of orthogonal polynomials can be generated using the Rodrigues' formula. \\

Consider a univariate random variable $U$ and let $P_k(U)$ denote a sequence of its orthogonal polynomials. Its Rodrigues' form yields: 
\begin{equation} \label{RF}
	P_k(U) = \frac{c_k}{w(U)} \frac{d^k}{dy^k} (w(U)Q(U)^n),
\end{equation}
where $c_n$ is a normalization factor, $w(U)$ is the weight function (proportional to the p.d.f of $U$) and $Q(U)$ is some polynomial function of $U$. For instance, the Hermite polynomials satisfy the Rodrigues form with $w(U)=\exp(-Y^2)$ and $Q(U) = 1$. Equation \eqref{RF} facilitates the construction of the orthogonal polynomials for random variables with weight function $w(U)$ analytically. Some examples of classical polynomials are shown in Table 4.  
\begin{table}[h]
	\centering
	\caption{Standard Orthogonal Polynomials and Their Properties}
	\begin{tabular}{llll}
\hline 
\hline 
		\textbf{Polynomial} & \textbf{Distribution} & \textbf{Weight} & \textbf{Rodrigues' Formula} \\
\hline 
\hline 
		Hermite &  $N(0,1)$ & $w(u)=\exp(-\frac{u^2}{2})$& $(-1)^n e^{u^2/2} \frac{d^n}{dy^n} e^{-u^2/2}$ \\
		Laguerre & $\text{Gamma}(\alpha+1,1)$ & $w(u)=u^\alpha \exp(-u)$& $\frac{1}{k!} u^{-\alpha} e^u \frac{d^k}{du^k} (u^{k+\alpha} e^{-k})$ \\
		Legendre & Unif$[-1,1]$ & $w(u)=1$ & $\frac{1}{2^k k!} \frac{d^k}{du^k} (u^2 - 1)^k$ \\
		Jacobi & $\text{Beta}(\alpha+1,\beta+1)$ & $w(u)=\frac{(1-u)^{-\alpha}}{(1+u)^{\beta}}$& $\frac{(-1)^k}{2^k k!} \frac{(1-u)^{-\alpha}}{(1+u)^{\beta}} \frac{d^k}{du^k} \left(\frac{(1-u)^{n+\alpha}}{(1+u)^{-n-\beta}}\right)$ \\
		Chebyshev & $\text{Beta}(\frac{1}{2},\frac{1}{2})$ & $w(u)=\frac{1}{\sqrt{1-u^2}}$& $\frac{(-1)^k}{2^{k-1} (k-1)!} (1 - u^2)^{\frac{1}{2}} \frac{d^k}{du^k} (1 - u^2)^{k- \frac{1}{2}}$\\
\hline 
\hline 
	\end{tabular}
\end{table}
Like the Hermite polynomial expansions, the classical orthogonal polynomials will also have their associated expansions provided trajectory of the process is square integrable with respect to their densities. For a collection of independent random variables $U_1,...,U_n$, the sequence of joint orthogonal polynomials are simply the product of the marginal orthogonal polynomials, that is:
\begin{equation}
	P_K(U_1,...,U_n) = \prod_{j=1}^n P_{k_j}(U_j),
\end{equation}
where $K=(k_1,...,k_n) \in \mathbb{N}^n$ and $k_j$ denotes the degree of nonlinearity of the $j$-th marginal orthogonal polynomial. \\

\noindent \textbf{Remark:} Orthogonal polynomials that are deemed not standard do not satisfy Rodrigues' formula and cannot be constructed in such a manner. A solution to this problem is to utilize a Gram-Schmidt procedure instead. However, it is important to note that such an approach relies on the existence of higher order moments. For distributions with fat tails for instance, the sequence of orthogonal polynomials may cease to exist. As an example, a t-distributed random variable has finite moments up to (but not including) its degree of freedom. This demonstrates that a variance based decomposition is generally limited to nonlinearities introduced by the dynamics of a process, and not suitable for nonlinearities arising from tail behaviour\footnote{This motivates an approach to decomposition analysis which relies on higher order moments, such as the case of the Forecast Relative Error Decompositions (FRED) introduced in \cite{GL2024}}.

\subsection{OFEVD Construction}

FEVD construction proceeds analogously to the HFEVD. The trajectory in \eqref{ftng} can be expressed as: 
\begin{equation}
	Y^*_{t+h} = g^{*(h)}(\underline{y_{t}},u_{t+1:t+h}) \approx \sum_{K \in \mathbb{R}^{nh}} C^{(h)}_{K}(\underline{y_t})P_K(u_{t+1:t+h}),
\end{equation}
where $P_K(u_{t+1:t+h}=\prod_{i=1}^h\prod_{j=1}^nP_k(u_{k,t+i})$. Then, the OFEVD is given by the following proposition: 
\begin{proposition}[Multivariate OHFEVD]
	Suppose the trajectory of an $n$-dimensional nonlinear SVAR($p$) is given by: 
	\begin{equation*}
		Y^*_{t+h} = g^{*(h)}(\underline{Y_{t}},u_{t+1:t+h}) ,
	\end{equation*}
	where each component of $g^{(h)}$ is square integrable and $u_{t+1:t+h}=(u_{1,t+1},...,u_{n,t+h})$ are independent structural innovations. Its variance covariance matrix conditional on the observed history $\underline{y_t}$ is given by: 
	%	\begin{equation}\label{fevd_quin_multi}
		%		\mathbb{V}[Y_{t+h}|\underline{y_t}] \approx \sum_{k=1}^{\infty}\sum_{\substack{K\in\mathbb{N}^{nh}  \\ K:\textbf{1}'K=k}}\left[C^{(h)}_K(\underline{y_t})C^{(h)}_K(\underline{y_t})'\prod_{i=1}^{nh}(k_{i}!)^2\right],
		%	\end{equation}
	%where $C^{(h)}_K(\underline{y_t}) = \mathbb{E}\left[ g^{(h)}(y_t,\varepsilon_{1,t+1:t+h},...,\varepsilon_{n,t+1:t+h})H_{K}(\varepsilon_{1,t+1:t+h},...,\varepsilon_{n,t+1:t+h})\right]/\prod_{i=1}^{nh}(k_{i}!)$ and $ \textbf{1}'K=k_1+...+k_{nh}$.
	\begin{equation}\label{ope}
		\mathbb{V}[Y^*_{t+h}|\underline{y_t}] \approx \sum_{K \in \mathcal{K}}\left[C^{(h)}_K(\underline{y_t})C^{(h)}_K(\underline{y_t})'\prod_{i=1}^{h}\prod_{j=1}^{n}\mathbb{V}[P_{k_{i,j}}(u_{k,t+i})]\right],
	\end{equation}
	where $\mathcal{K} = \left\{\mathbb{N}^{nh} \ \backslash \ \textbf{0}\right\}$ and  $C^{(h)}_K(\underline{y_t}) = \mathbb{E}\left[Y^*_{t+h}P_{K}(u_{t+1:t+h})\right]/\prod_{i=1}^{h}\prod_{j=n}^{nh}\mathbb{V}[P_{k}(u_{k,t+i})]$, where $P_{k}(u_{k,t+i})$ denotes the marginal orthogonal polynomial for innovation $u_{k,t+i})$ and $P_{K}(u_{t+1:t+h})$ is the joint orthogonal polynomial with degree characterized by $K \in \mathcal{K}$. 
\end{proposition}

To understand the differences between the OFEVD performed directly on non-Gaussian innovations versus the HFEVD on their standardized versions, consider a strong linear AR(1) process with independent innovations $u_t\sim t(5)$: 
\begin{equation}
	Y_{t} = aY_{t-1} + u_t.
\end{equation}
The trajectory at horizon 2 is given by:
\begin{equation}
	Y^*_{t+2} = a^2Y_t + au_{t+1}+u_{t+2}.
\end{equation}
The distribution on the innovation is characterized by fatter tails compared to the standard normal. Moreover, it does not have a valid Rodridgues' form, so the sequence of orthogonal polynomials cannot be generated by the formula in \eqref{RF}. As an alternative, the sequence can be generated by utilizing the Gram-Schmidt procedure for $u_t$\footnote{Only up to degree 4, since the Gram-Schmidt procedure for the 5th orthogonal polynomial would depend on $\mathbb{E}[U^5]$ which does not exist for the distribution for t(5) innovations. This showcases again the limitation of the orthogonal approach, since for fat-tailed distributions, the higher order terms may not exist even if the variance itself does.}. In this simple example, it suffices only to consider the first order orthogonal polynomials, since there are only first order terms of $u_{t+1}$ and $u_{t+2}$. In particular, the following terms can be computed from the Orthogonal FEVD:
\begin{equation}
	\begin{split}
		\left\{C_{(1,0)}^{(h)}(\underline{y_t})/\mathbb{V}[P_{1}(u_{t+1})]\right\}^2& = \mathbb{E}\left[Y^*_{t+2}u_{t+1}\right]^2=\left(\frac{5}{3}a\right)^2 \\ 
		\left\{C_{(0,1)}^{(h)}(\underline{y_t})/\mathbb{V}[P_{1}(u_{t+2})]\right\}^2& = \mathbb{E}\left[Y^*_{t+2}u_{t+2}\right]^2=\left(\frac{5}{3}\right)^2 \\ 
		\mathbb{V}\left[Y^*_{t+2}|Y_t\right] &= \mathbb{V}\left[au_{t+1}+u_{t+2}\right] = \left(\frac{5}{3}a\right)^2 + \left(\frac{5}{3}\right)^2
	\end{split}
\end{equation}
Hence there are two contributions to the variance of $Y_{t+2}^*$: the linear contribution of $u_{t+1}$ and the linear contribution of $u_{t+2}$. On the other hand, the main text suggests a normalization procedure where the errors $u_{t+1}$ and $u_{t+2}$ are transformed into Gaussian. This would yield a trajectory: 
\begin{equation}
		Y^*_{t+2} = a^2Y_t + a\Phi^{-1}(F_u(u_{t+1}))+\Phi^{-1}(F_u(u_{t+2})).
\end{equation}
Clearly, the HFEVD on Gaussian errors in this case would yield many terms including higher-order effects due to the nonlinearity arising from the function $\Phi^{-1}(F(\cdot))$. Ultimately, the choice of whether to use the OFEVD or HFEVD depends on the context of research and the views of the practitioner. However, it is clear that the OFEVD in this example would interpret the model as having only linear effects, whereas the HFEVD will interpret the model as having nonlinearities arising from the fatter tails of the t(5) distribution. In other words, nonlinearities can arise not only from the dynamics of the model (i.e. the relationship between $Y_{t}$ and its lags) but also from non-Gaussianity of the innovations. Moreover, the next subsection will show that the terms of the OFEVD no longer coincide with the impact multipliers of the EIRF when non-Gaussianity is present. This further complicates the interpretation of the OFEVD.

\subsection{FEVD and EIRF for Non-Gaussian Innovations}

The terms of the FEVD no longer coincide with the impact multipliers of the EIRF. It suffices to discuss the case of standard polynomials since the proofs in Appendices A.2 and A.3 depend on the existence of the Rodrigues' formula for Hermite polynomials (i.e. Equations \eqref{RFhu} and \eqref{RFhm}). Consider a simplified version of the proof in Appendix A.2 for exposition for some random variable $Y$. The coefficients in Equation \eqref{ope} are generated by: 
\begin{equation}
	c_k \times \langle P_{k}(U),P_{k}(U)\rangle = \mathbb{E}\left[g(U)P_k(U)\right] = \int_A g(u)P_k(u)w(u)dy,
\end{equation}
where $w(y)$ is a function that is proportional to the density function of $U$ on the support $A$.  By the Rodrigues' formula we get:
\begin{equation} 
	P_k(U) = \frac{1}{w(u)} \frac{d^k}{dU^k} (w(U)Q(U)^k),
\end{equation}
which implies:
\begin{equation} \label{RF2}
	P_k(U)w(U) = \frac{d^k}{dU^k} (w(U)Q(U)^n),
\end{equation}
Thus: 
\begin{equation}
	\begin{split}
		\mathbb{E}\left[g(U)P_k(U)\right] & = \int_A g(u)P_k(u)w(u)dy \\
		& =\int_A g(u)\frac{d^k}{du^k}(w(u)Q(u)^k)dy. \\ 	
	\end{split}
\end{equation}
In this step, as in \eqref{a5}, the integration by parts formula must be applied $k$ number of times. In the special case where $Q(U)=1$ (i.e. the Hermite polynomials), a nice tractable expression leads to the desired link between hte terms of the HFEVD and the impact multipliers of the EIRF. However, $Q(U)$ is in general, not a constant term and depends on the value of $U$. For instance, $Q(U)=U$ for the case of Laguerre polynomials, and $Q(U)=1-U^2$ for the Legendre polynomials. Thus, there is a compounding of terms introduced by the product rule, leading to a disconnect between the terms of the OFEVD and the EIRF. Indeed, this shows that Gaussianity of innovations is a very special case and one that is suitable for nonlinear FEVD analysis. Moreover, this result calls into question how the terms of a nonlinear FEVD can be interpreted if it is constructed directly using a definition of nonlinear IRFs [e.g. the \cite{LN2016} GFEVD approach uses the GIRF as a basis for FEVD consrtuction].

\subsection{Numerical Implementation for OFEVD}

The numerical implementation for the OFEVD can be performed by modifying the algorithms discussed in Section 5.1 and 5.2. Consider the $n$ dimensional SVAR(p) process described in \eqref{NLAR_alt}:
\begin{equation*}
	y_t = g^*(\underline{y_{t-1}},u_t;\theta) \iff u_t(\theta) = g^{*-1}(\underline{y_t};\theta).
\end{equation*}
We assume that this model is well-specified with a true value $\theta_0$ of the parameter, and that $u_t = u_t(\theta_0)$ is i.i.d. with a distribution $F_0$ which is unknown to the practitioner and not necessarily Gaussian. 

\begin{enumerate}
	\item Estimate $\theta$ by minimizing a covariance distance, for instance, by using the Generalized Covariance (GCov) approach [\cite{GJ2023}] written on tranformations of $u_t(\theta)$ and $u_{t-1}(\theta)$. If the model is linear in i.i.d. errors $u_t$, then it is also possible to apply a Psuedo Maximum Likelihood approach [\cite{GMR2017}]. Then, compute the residuals: $\hat{u}_{t} = g^{*-1}(\underline{y_t},\hat{\theta}_T)$. 
	\item Simulate (bootstrap) the future trajectory $S$ times, where each simulation is given by:
	\begin{equation}
		y_{t+h}^s = \hat{g}_T^{(h)}(\underline{y_t},u^s_{t+1},...,u^s_{t+h}),
	\end{equation} 
	and the $u^s_t$ are independently drawn from the sample distribution of $\hat{u}_{t}$, $t=1,...,T$.
	The value $\underline{y_t}$ is chosen either based on the practitioner's interest, on a particular historical profile, or by taking the last observations available in the data set. 
	\item Generate the seqeuence of orthogonal polynomials for the innovations innovations $u^s_{t+i}$ by applying the Gram-Schmidt procedure discussed in Appendix D.1.2. 
	\item Construct the relevant joint orthogonal polynomials of interest for all $K=(k_1,...,k_{nh})$ in the set $\Omega \subset \mathbb{N}^{nh}$ of interest.
	\item Compute the coefficients:
	\begin{equation}
		C^{(h)}_K(\underline{y_t}) = \mathbb{E}\left[Y^{*s}_{t+h}P_{K}(u^s_{t+1:t+h})\right]/\prod_{i=1}^{h}\prod_{j=1}^{n}\mathbb{V}[P_{k}(u^s_{k,t+i})]
	\end{equation}
	\item Compute the variance contribution for the terms of interest:
	\begin{equation}
		\sum_{\substack{K\in \Omega}}\left[C^{(h)}_K(\underline{y_t})C^{(h)}_K(\underline{y_t})'\prod_{i=1}^{h}\prod_{j=1}^{n}\mathbb{V}[P_{k}(u_{k,t+i})]^2\right].
	\end{equation}
\end{enumerate}

\end{document}